\documentclass[a4paper,11pt]{article}
\usepackage{jcappub}
\usepackage[T1]{fontenc}
\usepackage{graphicx}
\graphicspath{{figures/}}
\usepackage{amsmath}
\usepackage{amsfonts}
\usepackage{amssymb}
\usepackage{booktabs}
\usepackage{color}
\usepackage{natbib}

\newcommand{\kb}{\mathbf{k}}
\newcommand{\pb}{\mathbf{p}}
\newcommand{\qb}{\mathbf{q}}
\newcommand{\xbar}{\bar{x}}
\newcommand{\ubar}{\bar{u}}
\newcommand{\vbar}{\bar{v}}
\newcommand{\dd}{\mathrm{d}}
\newcommand{\cH}{\mathcal{H}}
\newcommand{\Lam}{\Lambda}
\newcommand{\Fsf}{\mathcal{F}}
\newcommand{\Jsf}{\mathcal{J}}
\newcommand{\Msf}{\mathcal{M}}
\newcommand{\Psf}{\mathcal{P}}
\newcommand{\Ssf}{\mathcal{S}}
\newcommand{\Pb}{\mathbb{P}}
\newcommand{\tk}{\tilde{k}}

\title{\boldmath Third-order scalar induced gravitational waves during a matter-dominated era: secular growth and implications for primordial black hole domination}

\author[a,1]{Jing-Zhi Zhou,\note{Corresponding author.}}

\affiliation[a]{Department of Mathematics and Physics, Huai'an University, Meicheng East Road 1, Huai'an, Jiangsu 223200, China}

\emailAdd{zhoujingzhi@hau.edu.cn}

\abstract{We investigate scalar induced gravitational waves (SIGWs) at third
order during a matter-dominated
(MD) era. We classify the sources into four parts---built from
three first-order scalars, and from a first-order scalar times the
second-order tensor, vector, and scalar perturbations---and obtain
all four kernel functions analytically in closed form. Unlike their
radiation-dominated counterparts, the MD kernel functions grow
secularly, translating into a rapid growth of the spectrum,
$\Omega^{(3)}_{\mathrm{GW}}\propto x^{6}$ versus
$\Omega^{(2)}_{\mathrm{GW}}\propto x^{2}$ at $x\equiv k\eta\gg1$.
We derive the third-order power spectrum as a five-fold integral for
a general primordial spectrum, and evaluate it numerically for a
monochromatic one, where all contributions from the second-order
vector source vanish identically. As an application, we consider the
early MD era driven by primordial black holes (PBHs) before their
evaporation. Requiring that the induced gravitational waves do not
backreact on the cosmological perturbations, $\int\mathrm{d}\ln k\,
\Omega^{\mathrm{tot}}_{\mathrm{GW}}<1$ then bounds the initial PBH
abundance as $\Omega_{\mathrm{PBH},f}\lesssim4.6\times10^{-7}\,
(m_{\mathrm{PBH}}/10^{7}\,\mathrm{g})^{-0.48}$, about three orders of
magnitude stronger than the second-order-only bound. Our results
indicate that higher-order SIGWs inevitably become nonperturbative
during a sufficiently long MD era, with important implications for
scenarios involving an early matter domination.}

\begin{document}

\maketitle
\flushbottom

\section{Introduction}\label{sec:intro}

The pulsar timing array (PTA) collaborations, including NANOGrav, the European PTA, the Parkes PTA and the Chinese PTA, have accumulated compelling evidence for a stochastic gravitational wave background (SGWB) in the nanohertz band \cite{NANOGrav:2023gor,EPTA:2023sfo,Reardon:2023gzh,Xu:2023wog}. Among the proposed cosmological interpretations, the scalar induced gravitational waves (SIGWs) are particularly attractive: the same enhanced small-scale curvature perturbations that source the SGWB may also seed the formation of primordial black holes (PBHs), providing a unified probe of the small-scale primordial power spectrum that is otherwise inaccessible to observations \cite{Sasaki:2018dmp,Carr:2020gox,Franciolini:2023pbf,Saito:2009jt}.

SIGWs are the tensor modes generated at second order in perturbation theory by the products of first-order scalar perturbations \cite{Mollerach:2003nq,Ananda:2006af,Baumann:2007zm,Saito:2008jc,Saito:2009jt,Bugaev:2010bb,Alabidi:2013lya,Malik:2008im}, where the connection with the PBH abundance was also established, and has since been extended in many directions, including non-Gaussian primordial perturbations \cite{Cai:2018dig,Ferrante:2022mui,Zhou:2025djn,Marriott-Best:2026ewq,Caporali:2026qhc,Iovino:2025cdy,Zeng:2025cer,Li:2025met,Inui:2024fgk,Pi:2024lsu,Perna:2024ehx,Li:2023xtl,Yuan:2023ofl,Ragavendra:2021qdu,Adshead:2021hnm,Yuan:2020iwf,Iovino:2024tyg}, gauge issues \cite{Domenech:2025ccu,Hwang:2017oxa,Tomikawa:2019tvi,Inomata:2019yww,Domenech:2020xin,Ali:2020sfw,DeLuca:2019ufz,Yuan:2019fwv,Lu:2020diy} and general cosmological backgrounds \cite{Domenech:2019quo,Liu:2023pau,Balaji:2023ehk,Domenech:2020kqm,Domenech:2024rks,Zhu:2023gmx,Franciolini:2026ner}; see Ref.~\cite{Domenech:2021ztg} for a review.

Of particular interest is the production of SIGWs during an early matter-dominated (MD) era. Such a phase is well motivated: ultralight PBHs with masses below $10^{9}\,\mathrm{g}$ evaporate before Big Bang nucleosynthesis and may temporarily dominate the energy density of the universe \cite{Sasaki:2018dmp,Carr:2020gox}, and other early MD episodes are also possible \cite{Silva-Malpartida:2024emu,Assadullahi:2009nf,Hajkarim:2019nbx,Pearce:2023kxp,Dalianis:2020gup,Pearce:2025ywc,Kumar:2024hsi,Dalianis:2024kjr,Harada:2022xjp,Das:2021wad,Zeng:2025ecx,Ling:2025hjx,Yoo:2026frk}. During matter domination the first-order scalar perturbations are frozen on all scales, and the induced tensor modes do not oscillate but grow linearly with the scale factor, so the resulting SIGW spectrum is qualitatively different from the radiation-dominated (RD) one. The second-order SIGWs during an early MD era, including the transition back to radiation domination, have been studied in detail \cite{Kohri:2018awv,Inomata:2018epa,Inomata:2019ivs,Inomata:2019zqy,Inomata:2020lmk,Domenech:2020ssp,Papanikolaou:2020qtd,Papanikolaou:2022chm}. In particular, Ref.~\cite{Papanikolaou:2020qtd} considered a universe filled with ultralight PBHs and derived strong constraints on the initial PBH abundance from the requirement that the second-order SIGWs do not backreact on the cosmological perturbations.

While the second-order SIGWs have been extensively investigated, the higher-order corrections are much less explored. The third-order SIGWs during the RD era were computed in Refs.~\cite{Zhou:2021vcw,Yuan:2021qgz,Yuan:2026krm,Chen:2022dah,Picard:2025bwq}, where the source was classified into four parts according to its perturbative content, and the phenomenological consequences for PBHs and for PTA observations were studied in Refs.~\cite{Chang:2023vjk,Wang:2023sij,Zhao:2026jax}. A systematic vertex-rule formalism for induced gravitational waves at arbitrary higher orders was developed in Ref.~\cite{Zhou:2024ncc}. However, the third-order SIGWs during a MD era have never been computed. Given that the second-order MD spectrum grows secularly as $\Omega^{(2)}_{\mathrm{GW}}\propto x^{2}$ with $x\equiv k\eta$, in sharp contrast to the RD case, it is urgent to clarify whether the higher-order corrections remain subdominant during a long MD era, or whether they grow even faster and eventually spoil the perturbative expansion.

In this paper we compute the third-order SIGWs during a MD era for the first time. We derive the third-order equation of motion, classify the source into four parts, and solve for all four kernel functions analytically in closed form. We find that the MD kernel functions grow secularly, $I^{(3)}\propto x^{2}$ at late times, which implies $\Omega^{(3)}_{\mathrm{GW}}\propto x^{6}$, growing as $x^{4}$ relative to the second-order spectrum. We derive the master formula for the third-order power spectrum as a five-fold integral for a general primordial spectrum, and evaluate it for a monochromatic primordial spectrum, for which we prove that all contributions involving the second-order vector source vanish identically. As an application, we revisit the PBH-dominated early universe of Ref.~\cite{Papanikolaou:2020qtd} and show that the secular growth makes the third-order spectrum dominate the total SIGW energy density at evaporation; the backreaction criterion applied to the total spectrum then strengthens the bound on the initial PBH abundance by about three orders of magnitude.

This paper is organized as follows. In section~\ref{sec:eom} we derive the equation of motion of the third-order SIGWs during matter domination and classify the source terms. In section~\ref{sec:kernel} we present the MD Green function and the second-order and third-order kernel functions in closed form, and discuss their secular growth. In section~\ref{sec:power} we derive the master formula for the third-order power spectrum and the energy density spectrum. In section~\ref{sec:mono} we evaluate the spectrum for a monochromatic primordial spectrum and compare the second-order and third-order results. In section~\ref{sec:pbh} we apply our results to the PBH-dominated early universe. We summarize and discuss the implications in section~\ref{sec:summary}. Technical details are collected in the appendices.

\section{Equation of motion}\label{sec:eom}
In this section, we derive and analyze the equation of motion for third-order SIGWs in the MD era. The $n$th-order perturbed metric in the flat Friedmann-Lemaitre-Robertson-Walker (FLRW) spacetime in Newtonian gauge is given by
\begin{equation}\label{eq:metric}
	\dd s^{2}=a^{2}(\eta)\Big[-\big(1+2\phi\big)\dd\eta^{2}
	+2V_{i}\,\dd\eta\,\dd x^{i}
	+\Big((1-2\psi)\delta_{ij}+h_{ij}\Big)\dd x^{i}\dd x^{j}\Big]\ ,
\end{equation}
where $\phi=\phi^{(1)}+\frac{1}{2}\phi^{(2)}+\cdots$ and $\psi=\psi^{(1)}+\frac{1}{2}\psi^{(2)}+\cdots$ denote scalar perturbations,  $V_{i}=\frac{1}{2}V^{(2)}_{i}+\frac{1}{6}V^{(3)}_{i}+\cdots$ denotes a vector perturbation, and $h_{ij}=\frac{1}{2}h^{(2)}_{ij}+\frac{1}{6}h^{(3)}_{ij}+\cdots$ denotes a tensor perturbation. In this study, we neglect the effects of potentially large-amplitude primordial vector and tensor perturbations on small scales. During matter domination the scale factor and the conformal Hubble parameter are
\begin{equation}\label{eq:MDbg}
	a(\eta)\propto\eta^{2}\,,\qquad \cH\equiv\frac{a'}{a}=\frac{2}{\eta}\,,
\end{equation}
and a prime denotes the derivative with respect to the conformal time $\eta$. The first-order scalar perturbations in momentum space are frozen on all scales
\begin{equation}\label{eq:phi1}
	\phi^{(1)}_{\pb}(\eta)=\Phi_{\pb}=\frac{3}{5}\,\zeta_{\pb}\,,
\end{equation}
where $\zeta_{\pb}$ is the primordial curvature perturbation. The equations of motion of the three kinds of second-order cosmological perturbations are given by \cite{Inomata:2020cck,Lu:2007cj,Kohri:2018awv}
\begin{eqnarray}
		&&\psi^{(2)}(\mathbf{x},\eta)-\phi^{(2)}(\mathbf{x},\eta)=-2 \Delta^{-1}\left(\partial^{r} \Delta^{-1} \partial^{s}-\frac{1}{2} \mathcal{T}^{rs}\right) \mathcal{S}^{(2)}_{rs}(\mathbf{x},\eta) \ , \label{eq:s2eom1}\\
		&&\psi^{(2)''}(\mathbf{x},\eta)+3\mathcal{H}  \psi^{(2)'}(\mathbf{x},\eta)-\frac{5}{6} \Delta \psi^{(2)}(\mathbf{x},\eta)+\mathcal{H}  \phi^{(2)'}(\mathbf{x},\eta)+\frac{1}{2} \Delta \phi^{(2)}(\mathbf{x},\eta)\nonumber\\
		&&~~~~~~~=-\frac{1}{2} \mathcal{T}^{rs} \mathcal{S}^{(2)}_{rs}(\mathbf{x},\eta) \ , \label{eq:S2eom2}\\
        &&V_{l}^{(2)'}(\mathbf{x},\eta)+2 \mathcal{H} V_{l}^{(2)}(\mathbf{x},\eta) =4 \Delta^{-1} \mathcal{T}_{l}^{r} \partial^{s} \mathcal{S}^{(2)}_{rs}(\mathbf{x},\eta) \ , \label{eq:V2eom}\\
        &&h_{lm}^{(2)''}(\mathbf{x},\eta)+2 \mathcal{H}  h_{lm}^{(2)'}(\mathbf{x},\eta)-\Delta h_{lm}^{(2)}(\mathbf{x},\eta)=-4 \Lambda_{lm}^{rs} \mathcal{S}^{(2)}_{rs}(\mathbf{x},\eta) \ , \label{eq:h2eom}
\end{eqnarray}
where $\Lambda_{rs}^{l m}=\mathcal{T}_{r}^{l} \mathcal{T}_{s}^{m}-\frac{1}{2} \mathcal{T}_{rs} \mathcal{T}^{l m}$ is the transverse and traceless operator, and $\mathcal{T}_{r}^{l}$ is defined as $\mathcal{T}_{r}^{l}=\delta_{r}^{l}-\partial^{l} \Delta^{-1} \partial_{r}$. The source term in Eq.~\eqref{eq:s2eom1}--Eq.~\eqref{eq:h2eom} is given by \cite{Chang:2020tji}
\begin{eqnarray}\label{eq:S2md}
	\mathcal{S}^{(2)}_{rs}(\mathbf{x},\eta)&=&\delta_{rs}\Big[(\phi^{(1)\prime})^{2}
	-2\Big(10\mathcal{H}\,\phi^{(1)}\phi^{(1)\prime}+(\phi^{(1)\prime})^{2}
	+2\phi^{(1)}\big(\phi^{(1)\prime\prime}+\Delta\phi^{(1)}\big)\Big)
	-3\,\partial_{b}\phi^{(1)}\partial^{b}\phi^{(1)}\Big]\nonumber\\
	&+&\frac{2}{3}\Big[-\frac{2}{\mathcal{H}^{2}}\,\partial_{r}\phi^{(1)\prime}
	\Big(\mathcal{H}\,\partial_{s}\phi^{(1)}+\partial_{s}\phi^{(1)\prime}\Big)
	+\partial_{r}\phi^{(1)}\Big(\partial_{s}\phi^{(1)}
	-\frac{2}{\mathcal{H}}\,\partial_{s}\phi^{(1)\prime}\Big)\nonumber\\	&+&6\,\phi^{(1)}\partial_{r}\partial_{s}\phi^{(1)}\Big]\ .
\end{eqnarray}
By solving Eq.~\eqref{eq:s2eom1}--Eq.~\eqref{eq:h2eom}, we obtain the formal expressions of the second-order perturbations in momentum space
\begin{eqnarray}
		\psi^{(2)}_{\mathbf{p}}(\eta)&=&\int \frac{\mathrm{d}^{3} q}{(2 \pi)^{3/2}}  I_{\psi}^{(2),MD}(|\mathbf{p}-\mathbf{q}|,|\mathbf{q}|,\eta)\zeta_{\mathbf{p}-\mathbf{q}} \zeta_{\mathbf{q}} \ , \label{eq:2psi0}\\
  \phi^{(2)}_{\mathbf{p}}(\eta)&=&\int \frac{\mathrm{d}^{3} q}{(2 \pi)^{3/2}}  I_{\phi}^{(2),MD}(|\mathbf{p}-\mathbf{q}|,|\mathbf{q}|,\eta)\zeta_{\mathbf{p}-\mathbf{q}} \zeta_{\mathbf{q}} \ , \label{eq:2phi0}\\
V^{\lambda,(2)}_{\mathbf{p}}(\eta)&=&\int \frac{\mathrm{d}^{3} q}{(2 \pi)^{3 / 2}}  \frac{ip^s e^{\lambda, r}(\mathbf{p})}{p^2} q_r q_s I_V^{(2),MD}(|\mathbf{p}-\mathbf{q}|,|\mathbf{q}|,\eta)\zeta_{\mathbf{p}-\mathbf{q}} \zeta_{\mathbf{q}} \ , \label{eq:2V0}\\
  	h^{\lambda,(2)}_{\mathbf{p}}(\eta)&=&\int \frac{\mathrm{d}^{3} q}{(2 \pi)^{3 / 2}}  \varepsilon^{\lambda, l m}(\mathbf{p})q_l q_m  I_{h}^{(2),MD}(|\mathbf{p}-\mathbf{q}|,|\mathbf{q}|,\eta)\zeta_{\mathbf{p}-\mathbf{q}}  \zeta_{\mathbf{q}}  \ , \label{eq:2h0}
\end{eqnarray}
where $I^{(2),MD}_i$ $(i=\psi,\phi,V,h)$ are the second-order kernel functions, whose closed-form expressions are given in section~\ref{sec:kernel2}. The superscript ``MD'' denotes the MD era.

By deriving the third-order cosmological perturbation equations and extracting their transverse-traceless part, we obtain the equation of motion for the third-order SIGWs
\begin{equation}\label{eq:h3eom}
	h_{i j}^{(3) \prime \prime}(\mathbf{x},\eta)+2 \mathcal{H} h_{i j}^{(3) \prime}(\mathbf{x},\eta)-\Delta h_{i j}^{(3)}(\mathbf{x},\eta)
	=-12\, \Lambda_{i j}^{l m}\, \mathcal{S}_{l m}^{(3)}(\mathbf{x},\eta) \ ,
\end{equation}
where the source term is decomposed into four parts according to its perturbative content,
\begin{equation}\label{eq:Sdec}
	\mathcal{S}_{l m}^{(3)}(\mathbf{x},\eta)=\mathcal{S}_{l m, 1}^{(3)}+\mathcal{S}_{l m, 2}^{(3)}
	+\mathcal{S}_{l m, 3}^{(3)}+\mathcal{S}_{l m, 4}^{(3)} \ .
\end{equation}
The first part is composed of three first-order scalar perturbations (we have set $\psi^{(1)}=\phi^{(1)}$),
\begin{align}
\label{eq:S1}
	\mathcal{S}_{l m, 1}^{(3)}&=-\frac{16}{3 \mathcal{H}}\,\phi^{(1)\prime} \partial_l \phi^{(1)} \partial_m \phi^{(1)}
	+\frac{8}{9 \mathcal{H}^2}\,\Delta \phi^{(1)} \partial_l \phi^{(1)} \partial_m \phi^{(1)}
	-\frac{4}{\mathcal{H}^2}\,\phi^{(1)\prime} \partial_l \phi^{(1)\prime} \partial_m \phi^{(1)}\nonumber\\
	&+\frac{8}{9 \mathcal{H}^3}\,\Delta \phi^{(1)} \partial_l \phi^{(1)\prime} \partial_m \phi^{(1)}
	-\frac{4}{\mathcal{H}^2}\,\phi^{(1)\prime} \partial_l \phi^{(1)} \partial_m \phi^{(1)\prime}
	+\frac{8}{9 \mathcal{H}^3}\,\Delta \phi^{(1)} \partial_l \phi^{(1)} \partial_m \phi^{(1)\prime}\nonumber\\
	&-\frac{16}{3 \mathcal{H}^2}\,\phi^{(1)} \partial_l \phi^{(1)\prime} \partial_m \phi^{(1)\prime}
	-\frac{8}{3 \mathcal{H}^3}\,\phi^{(1)\prime} \partial_l \phi^{(1)\prime} \partial_m \phi^{(1)\prime}
	+\frac{8}{9 \mathcal{H}^4}\,\Delta \phi^{(1)} \partial_l \phi^{(1)\prime} \partial_m \phi^{(1)\prime}\nonumber\\
    &+\frac{40}{3}\, \phi^{(1)} \partial_l \phi^{(1)} \partial_m \phi^{(1)}\ .
\end{align}
The second part is composed of the first-order scalar and the second-order tensor,
\begin{align}
\label{eq:S2}
	\mathcal{S}_{l m, 2}^{(3)}={}&-\frac{1}{2} h^{(2)\prime\prime}_{lm} \phi^{(1)}
	-\cH\,h^{(2)\prime}_{lm} \phi^{(1)}+4\cH\, h^{(2)}_{lm} \phi^{(1)\prime}
	+\frac{3}{2} h^{(2)}_{lm} \phi^{(1)\prime\prime}
	-\frac{1}{2} \phi^{(1)} \Delta h^{(2)}_{lm}\nonumber\\
	&+\frac{1}{2} h^{(2)b}_{\quad m} \partial_b \partial_l \phi^{(1)}
	+\frac{1}{2} h^{(2)b}_{\quad l} \partial_b \partial_m \phi^{(1)}
	-\partial_b h^{(2)}_{lm} \partial^b \phi^{(1)}
	+\frac{1}{2} \partial^b \phi^{(1)} \partial_l h^{(2)}_{mb} \nonumber\\
	&+\frac{1}{2} \partial^b \phi^{(1)} \partial_m h^{(2)}_{lb}-\frac{1}{2} h^{(2)}_{lm} \Delta \phi^{(1)}\ .
\end{align}
The third part is composed of the first-order scalar and the second-order vector,
\begin{align}
\label{eq:S3}
	\mathcal{S}_{l m, 3}^{(3)}={}&\mathcal{H} \phi^{(1)} \partial_l V^{(2)}_{m}
	+\frac{1}{2} \phi^{(1)\prime} \partial_l V^{(2)}_{m}
	+\frac{1}{2} \phi^{(1)} \partial_l V^{(2)\prime}_{m}
	-\frac{1}{2} V^{(2)\prime}_{m} \partial_l \phi^{(1)}
	-\cH\,V^{(2)}_{m} \partial_l \phi^{(1)}\nonumber\\
	&+\frac{1}{6 \mathcal{H}}\,\Delta V^{(2)}_{m} \partial_l \phi^{(1)}
	-\frac{1}{2} V^{(2)}_{m} \partial_l \phi^{(1)\prime}
	+\frac{1}{6 \mathcal{H}^2}\,\Delta V^{(2)}_{m} \partial_l \phi^{(1)\prime}
	+\big(l\leftrightarrow m\big)\ ,
\end{align}
where $(l\leftrightarrow m)$ denotes the eight terms obtained by interchanging the free indices. The fourth part is composed of the first-order scalar and the second-order scalars,
\begin{align}
\label{eq:S4}
	\mathcal{S}_{l m, 4}^{(3)}={}&-\frac{2}{3} \partial_l \phi^{(2)} \partial_m \phi^{(1)}
	+\partial_l \psi^{(2)} \partial_m \phi^{(1)}
	-\frac{2}{3 \mathcal{H}}\partial_l \psi^{(2)\prime} \partial_m \phi^{(1)}
	-\frac{2}{3 \mathcal{H}}\partial_l \phi^{(2)} \partial_m \phi^{(1)\prime}\nonumber\\
	&-\frac{2}{3} \partial_l \phi^{(1)} \partial_m \phi^{(2)}
	-\frac{2}{3 \mathcal{H}}\partial_l \phi^{(1)\prime} \partial_m \phi^{(2)}
	+\partial_l \phi^{(1)} \partial_m \psi^{(2)}
	-\frac{2}{3 \mathcal{H}}\partial_l \phi^{(1)} \partial_m \psi^{(2)\prime}\nonumber\\
	&+\phi^{(2)} \partial_l\partial_m \phi^{(1)}
	+\psi^{(2)} \partial_l\partial_m \phi^{(1)}
	+\phi^{(1)} \partial_l\partial_m \phi^{(2)}
	+\phi^{(1)} \partial_l\partial_m \psi^{(2)}\nonumber\\
    &-\frac{2}{3 \mathcal{H}^2}\partial_l \psi^{(2)\prime} \partial_m \phi^{(1)\prime} -\frac{2}{3 \mathcal{H}^2}\partial_l \phi^{(1)\prime} \partial_m \psi^{(2)\prime} \ .
\end{align}
Terms involving the third-order scalar or vector perturbations are annihilated by the transverse and traceless operator  $\Lam_{ij}^{lm}$ and do not enter Eq.~\eqref{eq:h3eom}. The above equations are obtained with the same tensor conventions as in the RD analysis of Ref.~\cite{Zhou:2021vcw}. By solving Eq.~\eqref{eq:h3eom} in momentum space, we obtain the formal expressions of the third-order SIGWs 
\begin{eqnarray}\label{eq:3h}
h^{\lambda,(3)}_{\mathbf{k}}(\eta)=
h_{ 1,\mathbf{k}}^{\lambda,(3)}(\eta)+
h_{ 2,\mathbf{k}}^{\lambda,(3)}(\eta)+
h_{ 3,\mathbf{k}}^{\lambda,(3)}(\eta)+
h_{ 4,\mathbf{k}}^{\lambda,(3)}(\eta) \  ,
\end{eqnarray}
where
\begin{eqnarray}
h_{1,\mathbf{k}}^{\lambda,(3)}(\eta)&=&\int \frac{\mathrm{d}^3 p}{(2 \pi)^{3 / 2}} \int \frac{\mathrm{d}^3 q}{(2 \pi)^{3 / 2}} \varepsilon^{\lambda, l m}(\mathbf{k}) q_m \left(p_l-q_l\right) I_{1}^{(3),MD} \zeta_{\mathbf{k}-\mathbf{p}} \zeta_{\mathbf{p}-\mathbf{q}} \zeta_{\mathbf{q}}\ , \label{eq:3h1}\\
h_{2,\mathbf{k}}^{\lambda,(3)}(\eta)&=&\int \frac{\mathrm{d}^3 p}{(2 \pi)^{3 / 2}} \int \frac{\mathrm{d}^3 q}{(2 \pi)^{3 / 2}} \varepsilon^{\lambda, l m}(\mathbf{k}) \Lambda_{l m}^{r s}(\mathbf{p}) q_r q_s  I_{2}^{(3),MD} \zeta_{\mathbf{k}-\mathbf{p}} \zeta_{\mathbf{p}-\mathbf{q}} \zeta_{\mathbf{q}}\ , \label{eq:3h2}\\
h_{ 3,\mathbf{k}}^{\lambda,(3)}(\eta)&=&\int \frac{\mathrm{d}^3 p}{(2 \pi)^{3 / 2}} \int \frac{\mathrm{d}^3 q}{(2 \pi)^{3 / 2}} \varepsilon^{\lambda, l m}(\mathbf{k}) \left(\mathcal{T}_{m}^r(\mathbf{p}) p_{l}+\mathcal{T}_{l}^r(\mathbf{p}) p_{m}\right)  \frac{p^s}{ p^2} q_r q_s \nonumber\\
&\times& I_{3}^{(3),MD} \zeta_{\mathbf{k}-\mathbf{p}} \zeta_{\mathbf{p}-\mathbf{q}} \zeta_{\mathbf{q}}\ , \label{eq:3h3}\\
h_{ 4,\mathbf{k}}^{\lambda,(3)}(\eta)&=&\int \frac{\mathrm{d}^3 p}{(2 \pi)^{3 / 2}} \int \frac{\mathrm{d}^3 q}{(2 \pi)^{3 / 2}} \varepsilon^{\lambda, l m}(\mathbf{k}) p_l p_m   I_{4}^{(3),MD} \zeta_{\mathbf{k}-\mathbf{p}} \zeta_{\mathbf{p}-\mathbf{q}} \zeta_{\mathbf{q}} \  . \label{eq:3h4}
\end{eqnarray}
Here $\Lambda_{lm}^{rs}(\pb)$ and $\mathcal{T}_{m}^{r}(\pb)$ are the momentum-space counterparts of the operators defined below Eq.~\eqref{eq:h2eom}. In particular, the formal structures of the four sources are identical to the RD case, and only the time-dependent kernel functions $I_i^{(3),MD}$ $(i=1,2,3,4)$ are modified, whose closed-form expressions are given in section~\ref{sec:kernel3}.

\section{Kernel functions}\label{sec:kernel}

In this section we solve the equations of motion derived in
section~\ref{sec:eom} analytically. We first construct the MD Green
function in section~\ref{sec:green}, then present the closed-form MD
kernel functions of the second-order perturbations entering
Eq.~\eqref{eq:2psi0}--Eq.~\eqref{eq:2h0} in section~\ref{sec:kernel2},
and finally derive the four third-order kernel functions entering
Eq.~\eqref{eq:3h1}--Eq.~\eqref{eq:3h4} in section~\ref{sec:kernel3},
whose secular growth is the origin of all the qualitatively new
results of this paper.

\subsection{Green function}\label{sec:green}

Both the second-order tensor equation \eqref{eq:h2eom} and the
third-order one \eqref{eq:h3eom} share the same differential operator
on the left-hand side, so we first construct the common MD Green
function. We introduce the dimensionless variables
\begin{equation}\label{eq:vars}
	x=k\eta\,,\qquad
	u=\frac{|\kb-\pb|}{k}\,,\qquad v=\frac{p}{k}\,,\qquad
	y=p\eta=vx\,,\qquad
	\ubar=\frac{|\pb-\qb|}{p}\,,\qquad \vbar=\frac{q}{p}\,,
\end{equation}
where $\kb$ is the external momentum of the induced tensor and
$\pb,\qb$ are the loop momenta of the sources. The equation of
motion of the third-order SIGWs in momentum space is then given by
\begin{equation}\label{eq:h3mode}
	h^{\lambda,(3)\prime\prime}_{\kb}+2\cH h^{\lambda,(3)\prime}_{\kb}
	+k^{2}h^{\lambda,(3)}_{\kb}=12\,\Ssf^{\lambda}_{\kb}\,,
\end{equation}
with $\Ssf^{\lambda}\equiv-\varepsilon^{\lambda,lm}(\kb)\Ssf^{(3)}_{lm}$ and $\varepsilon^{\lambda,lm}(\kb)$ the polarization tensor. The solution to Eq.~\eqref{eq:h3mode} can then be written in terms of the MD Green function
\begin{equation}\label{eq:GMD}
	k\,G^{MD}_{k}(x,\xbar)
	=\frac{(1+x\xbar)\sin(x-\xbar)-(x-\xbar)\cos(x-\xbar)}{x\,\xbar}\,,
	\qquad \xbar=k\bar\eta\,.
\end{equation}
Every third-order kernel function is defined with the source coefficient $12$ of Eq.~\eqref{eq:h3mode} factored out uniformly:
\begin{equation}\label{eq:kernelInt}
	I^{(3)}_{i}(x)=\frac{12}{k^{2}}\int_{0}^{x}\dd\xbar
	\left(\frac{\xbar}{x}\right)^{2}k\,G^{MD}_{k}(x,\xbar)\,
	f^{(3),MD}_{i}(\xbar)\,  , \, \, \, (i=1,2,3,4) \,,
\end{equation}
where $f^{(3),MD}_{i}$ is the time-dependent coefficient of the corresponding momentum-space source. Equivalently, each kernel function obeys
\begin{equation}\label{eq:kernelODE}
	\partial_{x}^{2}I^{(3)}_{i}+\frac{4}{x}\,\partial_{x}I^{(3)}_{i}+I^{(3)}_{i}
	=\frac{12}{k^{2}}\,f^{(3),MD}_{i}\,,\qquad
	I^{(3)}_{i}(0)=\partial_{x}I^{(3)}_{i}(0)=0\,.
\end{equation}
The same Green function with the coefficient $12$ replaced by $4$
solves the second-order tensor equation \eqref{eq:h2eom}.

\subsection{Second-order MD kernel functions}\label{sec:kernel2}

The second-order cosmological perturbations are determined by the second-order perturbation equations and subsequently enter the third-order sources in Eq.~\eqref{eq:S2}--Eq.~\eqref{eq:S4}; we therefore solve them first. During the MD era, the first-order scalar perturbations are frozen, and the kernel function of the second-order tensor perturbation is given by \cite{Kohri:2018awv}
\begin{equation}\label{eq:Ih2}
	I^{(2),MD}_{h}(y)=\frac{24}{5p^{2}}\,\Fsf(y)=\frac{24}{5p^{2}}\,\left(\frac{y^{3}+3y\cos y-3\sin y}{y^{3}}\right)\,,
\end{equation}
where $y=p\eta=vx$ in the notation of Eq.~\eqref{eq:vars}. The second-order vector equation \eqref{eq:V2eom} has the same
time-independent source, and its solution entering Eq.~\eqref{eq:2V0}
grows linearly in time,
\begin{equation}\label{eq:IV2}
	I^{(2),MD}_{V}(\eta)=\frac{24}{25}\,\eta\,.
\end{equation}
It should be noted that the factor of $(3/5)^2$ arising from the primordial curvature perturbation in Eq.~\eqref{eq:phi1} has been absorbed into the second-order kernel functions. Finally, the two second-order scalar perturbations obtained from
Eq.~\eqref{eq:s2eom1} and Eq.~\eqref{eq:S2eom2} enter
Eq.~\eqref{eq:2psi0} and Eq.~\eqref{eq:2phi0} with the kernel functions \cite{Inomata:2020cck}
\begin{equation}\label{eq:Ipsi2}
	\begin{aligned}
		I^{(2),MD}_{\psi}(\ubar,\vbar,y)={}
		\frac{9}{25}&\Big\{\frac{1}{3}\Big[2(\ubar^{2}+\vbar^{2})-3(\ubar^{2}-\vbar^{2})^{2}
		+5-10a_{\mathrm{NL}}\Big]\\
		&-\frac{y^{2}}{84}\Big[2+3(\ubar^{2}+\vbar^{2})
		-5(\ubar^{2}-\vbar^{2})^{2}\Big]\Big\}\,,
	\end{aligned}
\end{equation}
\begin{equation}\label{eq:Iphi2}
	I^{(2),MD}_{\phi}(\ubar,\vbar,y)=I^{(2),MD}_{\psi}(\ubar,\vbar,y)
	+\frac{3}{50}\Big[19-10(\ubar^{2}+\vbar^{2})+15(\ubar^{2}-\vbar^{2})^{2}\Big]\,.
\end{equation}
The constant term of Eq.~\eqref{eq:Ipsi2} encodes the super-horizon initial value of the second-order curvature perturbation, parameterized as $\zeta^{(2)}=2a_{\mathrm{NL}}(\zeta^{(1)})^{2}$ \cite{Inomata:2020cck}. Gaussian primordial perturbations correspond to $a_{\mathrm{NL}}=1$ \cite{Bartolo:2004ty,Bartolo:2005xa}, which we adopt throughout, consistently with the Wick expansion of the six-point scalar correlator in section~\ref{sec:power}. Both scalar kernel functions grow as $y^{2}$ at
late times, a growth inherited by the third-order source in Eq.~\eqref{eq:S4}.

\subsection{Third-order MD kernel functions}\label{sec:kernel3}

We finally derive the four third-order kernel functions entering
Eq.~\eqref{eq:3h1}--Eq.~\eqref{eq:3h4}. Inserting the classified
sources Eq.~\eqref{eq:S1}--Eq.~\eqref{eq:S4}, with the second-order
perturbations replaced by the solutions of section~\ref{sec:kernel2},
into the source term of Eq.~\eqref{eq:h3mode}, the corresponding source functions in
Eq.~\eqref{eq:kernelInt} read
    \begin{align}
        f^{(3),MD}_{1}(u,x)&=\frac{40}{3}-\frac{2}{9}\,u^{2}x^{2}\,, \label{eq:f1}\\
        f^{(3),MD}_{2}(v,x)&=\frac{20}{3}-\frac{20(1+v^{2})}{3v^{2}}\,\Fsf(vx)\,,\label{eq:f2}\\
        f^{(3),MD}_{3}(v,x)&=\frac{40}{3}+\frac{2}{9}\,v^{2}x^{2}\,,\label{eq:f3}\\
        f^{(3),MD}_{4}(v,\ubar,\vbar,x)&=C(\ubar,\vbar)
		-\frac{v^{2}x^{2}}{18}\,B(\ubar,\vbar)\,, \label{eq:f4}
    \end{align}
where
\begin{equation}
	\begin{aligned}
		B(\ubar,\vbar)=&2+3(\ubar^{2}+\vbar^{2})
		-5(\ubar^{2}-\vbar^{2})^{2}\,,\\
		C(\ubar,\vbar)=\frac{10}{3}&\left(
		\frac{2(\ubar^{2}+\vbar^{2})-3(\ubar^{2}-\vbar^{2})^{2}
		+5-10a_{\mathrm{NL}}}{3} \right. \\
		&\left.+\frac{19-10(\ubar^{2}+\vbar^{2})+15(\ubar^{2}-\vbar^{2})^{2}}{6}
		\right).
	\end{aligned}
\end{equation}
Here $f^{(3),MD}_{2}$ and $f^{(3),MD}_{3}$ inherit the MD second-order
tensor and vector kernel functions \eqref{eq:Ih2} and \eqref{eq:IV2},
while $B(\ubar,\vbar)$ and $C(\ubar,\vbar)$ inherit the second-order
scalar kernel functions \eqref{eq:Ipsi2} and \eqref{eq:Iphi2}. With the source functions at hand, the time integrals in
Eq.~\eqref{eq:kernelInt} are elementary. For convenience, we define the following two integrals
\begin{equation}\label{eq:J2}
	\Jsf_{2}(x)\equiv\int_{0}^{x}\dd\xbar
	\left(\frac{\xbar}{x}\right)^{2}kG^{MD}_{k}(x,\xbar)\,\xbar^{\,2}
	=x^{2}-10-\frac{30\cos x}{x^{2}}+\frac{30\sin x}{x^{3}}\,,
\end{equation}
and
\begin{equation}\label{eq:M0}
	\begin{aligned}
		\Msf_{0}(x,v)\equiv{}&\int_{0}^{x}\dd\xbar
		\left(\frac{\xbar}{x}\right)^{2}kG^{MD}_{k}(x,\xbar)\,\Fsf(v\xbar)\\
		={}&\frac{v^{3}(v^{2}-1)x^{3}-3vx\cos(vx)+3v^{5}(x\cos x-\sin x)
		+3\sin(vx)}{v^{3}(v^{2}-1)x^{3}}\,.
	\end{aligned}
\end{equation}
Then the third-order kernel functions entering the formal
solutions Eq.~\eqref{eq:3h1}--Eq.~\eqref{eq:3h4} are obtained in closed form:
\begin{equation}\label{eq:I1}
	I^{(3),MD}_{1}(u,x)=\left(\frac{3}{5}\right)^{3}\frac{12}{k^{2}}\left[
	\frac{40}{3}\,\Fsf(x)-\frac{2u^{2}}{9}\,\Jsf_{2}(x)\right],
\end{equation}
\begin{equation}\label{eq:I2}
	I^{(3),MD}_{2}(v,x)=\left(\frac{3}{5}\right)^{3}\frac{12}{k^{2}}\left[\frac{20}{3}\,\Fsf(x)
	-\frac{20(1+v^{2})}{3v^{2}}\,\Msf_{0}(x,v)\right],
\end{equation}
\begin{equation}\label{eq:I3}
	I^{(3),MD}_{3}(v,x)=\left(\frac{3}{5}\right)^{3}\frac{12}{k^{2}}\left[
	\frac{40}{3}\,\Fsf(x)+\frac{2v^{2}}{9}\,\Jsf_{2}(x)\right],
\end{equation}
\begin{equation}\label{eq:I4}
	I^{(3),MD}_{4}(v,\ubar,\vbar,x)=\left(\frac{3}{5}\right)^{3}\frac{12}{k^{2}}\left[
	C(\ubar,\vbar)\,\Fsf(x)-\frac{v^{2}}{18}\,B(\ubar,\vbar)\,
	\Jsf_{2}(x)\right].
\end{equation}

We close this section with the asymptotic behavior of the third-order kernel functions. At late times the secular growth $\Jsf_{2}(x)\propto x^{2}$ dominates:
\begin{equation}\label{eq:Ilate}
	\begin{aligned}
		I^{(3),MD}_{1}&\simeq-\left(\frac{3}{5}\right)^{3}\frac{8u^{2}x^{2}}{3k^{2}}\,,\qquad
		I^{(3),MD}_{2}\simeq-\left(\frac{3}{5}\right)^{3}\frac{80}{k^{2}v^{2}}\,,\\
		I^{(3),MD}_{3}&\simeq\left(\frac{3}{5}\right)^{3}\frac{8v^{2}x^{2}}{3k^{2}}\,,\qquad
		I^{(3),MD}_{4}\simeq-\left(\frac{3}{5}\right)^{3}\frac{2v^{2}B(\ubar,\vbar)}{3}\,\frac{x^{2}}{k^{2}}\,,
	\end{aligned}
\end{equation}
so three of the four kernel functions grow as $x^{2}$, while
$I^{(3),MD}_{2}$ asymptotes to a constant. This is in sharp contrast
with the second-order tensor kernel function \eqref{eq:Ih2}, which
asymptotes to a constant since $\Fsf (x)\to1$ at late times. The secular growth has a
different origin in each source function. In $f^{(3),MD}_{1}$ the
first-order scalars are frozen, so the time-dependent piece comes
solely from the Laplacian term of Eq.~\eqref{eq:S1}: its coefficient
$8/(9\cH^{2})$ grows as $\eta^{2}$ during the MD era and turns into the
$u^{2}x^{2}$ term. In $f^{(3),MD}_{3}$ and $f^{(3),MD}_{4}$ the
growth is inherited from the second-order perturbations in the
sources: the vector perturbation \eqref{eq:IV2} grows linearly in
$\eta$, while the scalar kernel functions \eqref{eq:Ipsi2} and
\eqref{eq:Iphi2} grow as $y^{2}$. In contrast, the second-order
tensor kernel function \eqref{eq:Ih2} entering $f^{(3),MD}_{2}$
asymptotes to a constant since $\Fsf(x)\to1$ at late times, so
$I^{(3),MD}_{2}$ does not grow. None of these mechanisms operates
during radiation domination, where the first-order scalars oscillate
and decay inside the horizon and the second-order perturbations do
not grow.

\section{Energy density spectrum}\label{sec:power}

In this section we derive the master formula for the third-order SIGW
power spectrum during matter domination. We set up the two-point
function and the Wick expansion of the six-point scalar correlator in
section~\ref{sec:wick}, reduce the integration measure to a five-fold
form in section~\ref{sec:measure}, present the master formula and an
exact symmetry of the cross spectra in section~\ref{sec:masterf}, and
fix the normalization of the energy density spectrum in
section~\ref{sec:omega}.

\subsection{Formal expression}\label{sec:wick}

The power spectrum of the third-order SIGW is defined by
\begin{equation}\label{eq:2pt}
	\langle h^{\lambda,(3)}(\eta,\kb)h^{\lambda',(3)}(\eta,\kb')\rangle
	=\delta^{\lambda\lambda'}\delta(\kb+\kb')\,
	\frac{2\pi^{2}}{k^{3}}\,\Psf^{(3)}_{h}(\eta,k)\,,\qquad
	\Psf^{(3)}_{h}=\sum_{i,j=1}^{4}\Psf^{ij}_{h}\,.
\end{equation}
Substituting Eq.~\eqref{eq:3h1}--Eq.~\eqref{eq:3h4} gives the twelve-dimensional integral
\begin{align}\label{eq:Ppij}
	\Psf^{ij}_{h}(\eta,k)&=\frac{k^{3}(2\pi^{2})^{2}}{2}
	\int\frac{\dd^{3}p\,\dd^{3}q\,\dd^{3}p'\,\dd^{3}q'}{(2\pi)^{6}}\,
	\Pb^{ij}(\kb,\pb,\qb;\pb',\qb')\nonumber\\
	&\times\mathcal{C}(\kb,\kb',\pb,\qb,\pb',\qb')
	I^{(3),MD}_{i}\,I^{(3),MD\prime}_{j}\,.
\end{align}
The integrand of the third-order SIGW power spectrum in the MD era, as given in Eq.~\eqref{eq:Ppij}, consists of three distinct parts: the momentum polynomial $\Pb^{ij}$, which arises from contractions of the polarization tensors with the momentum; the factor $\mathcal{C}$, which originates from the six-point correlation function of the primordial curvature perturbation; and the product of two third-order SIGW kernel functions. In the following, we briefly analyze each of these ingredients.

The momentum polynomial $\Pb^{ij}(\kb,\pb,\qb;\pb',\qb')$ in Eq.~\eqref{eq:Ppij} is the contraction of the polarization
tensors with the momentum structures. The polarization sum obeys the
completeness relation \cite{Zhou:2021vcw}
\begin{equation}\label{eq:Lam}
	\delta^{\lambda\bar{\lambda}}\varepsilon^{\bar{\lambda},lm}(\kb)
	\varepsilon^{\bar{\lambda}}_{ij}(\kb)
	=\Lambda_{ij}^{lm}(\hat\kb)+A_{ij}^{lm}(\kb)\,,
\end{equation}
where the second term
\begin{equation}\label{eq:Aij}
	A_{ij}^{lm}(\kb)\equiv\frac{1}{\sqrt{2}}\left(\bar{e}_i(\kb)e_j(\kb)
	-e_i(\kb)\bar{e}_j(\kb) \right)\frac{1}{\sqrt{2}}
	\left(\bar{e}^m(\kb)e^l(\kb)-e^m(\kb)\bar{e}^l(\kb) \right)
\end{equation}
depends on the transverse polarization vectors $e_i(\kb)$ and
$\bar{e}_i(\kb)$. As shown by Eq.~\eqref{eq:Aij}, $A_{ij}^{lm}$ is
antisymmetric in both index pairs, so $A_{ij}^{lm}T_{lm}
=A_{ij}^{lm}T^{ij}=0$ for an arbitrary symmetric tensor $T_{lm}$.
Since the polarization tensors are symmetric in $l\leftrightarrow m$,
the momentum structures below can be taken symmetric without loss of
generality, and the momentum polynomial reduces to the contraction
with the transverse-traceless operator,
\begin{equation}\label{eq:Pijdef}
	\Pb^{ij}(\kb,\pb,\qb;\pb',\qb')
	=\Lam_{lm,l'm'}(\hat{\kb})\,\Ssf^{lm}_{i}(\pb,\qb)\,
	\Ssf^{l'm'}_{j}(\pb',\qb')\,,
\end{equation}
with the momentum structures
\begin{equation}\label{eq:structs}
	\begin{aligned}
		\Ssf^{lm}_{1}&=(p_{l}-q_{l})q_{m}\,,\qquad
		\Ssf^{lm}_{2}=\Lam_{lm,rs}(\pb)\,q_{r}q_{s}\,,\qquad
		\Ssf^{lm}_{4}=p_{l}p_{m}\,,\\
		\Ssf^{lm}_{3}&=\Big[P_{m}{}^{r}(\pb)\,p_{l}
		+P_{l}{}^{r}(\pb)\,p_{m}\Big]\frac{p^{s}}{p^{2}}\,q_{r}q_{s}
		=\frac{\pb\cdot\qb}{p^{2}}\Big[p_{l}\bar q_{m}+p_{m}\bar q_{l}\Big]\,,
		\qquad \bar q_{m}\equiv P_{m}{}^{r}(\pb)\,q_{r}\,,
	\end{aligned}
\end{equation}
where $P_{lm}(\pb)\equiv\delta_{lm}-\hat{p}_{l}\hat{p}_{m}$ is the
momentum-space form of the transverse operator $\mathcal{T}_{l}^{m}$
defined below Eq.~\eqref{eq:h2eom}. 

Furthermore, the factor $\mathcal{C}(\kb,\kb',\pb,\qb,\pb',\qb')$ in Eq.~\eqref{eq:Ppij} comes from the six-point function
$\langle\zeta_{\kb-\pb}\zeta_{\pb-\qb}\zeta_{\qb}
\zeta_{\kb'-\pb'}\zeta_{\pb'-\qb'}\zeta_{\qb'}\rangle$.
Wick's theorem gives $15$ pairings: $6$ cross contractions, in which
the three $\zeta$'s of $h^{\lambda,(3)}$ are paired with those of
$h^{\lambda',(3)}$, correspond to one-particle-irreducible loop
diagrams, while the remaining $9$ pairings are one-particle reducible.
The reducible pairings vanish identically once the azimuthal
integration of the loop momentum is performed, as proven rigorously
with the vertex rules in Ref.~\cite{Zhou:2024ncc}, so only the $6$
irreducible pairings contribute to the power spectrum for an arbitrary
primordial spectrum. Each of them yields the same product of spectra $P_{\zeta}(ku)\,P_{\zeta}(\ubar vk)\,P_{\zeta}(\vbar vk)$ and fixes $(\pb',\qb')$ to one of the six substitutions listed in table~\ref{tab:wick} of appendix~\ref{Ap:wick}, together with the
corresponding arguments $(u',v',\ubar',\vbar')$ of the kernel functions. 

Moreover, a key property of the MD kernel functions, used repeatedly below, is their
exact symmetry under the exchange of the last two scalar-momentum
arguments,
\begin{equation}\label{eq:MDsym}
	I^{(3),MD}_{j}(u,v,\ubar,\vbar,x)
	=I^{(3),MD}_{j}(u,v,\vbar,\ubar,x)\,,
\end{equation}
which holds because $B$, $C$ depend on $\ubar,\vbar$ only through the
combinations $\ubar^{2}+\vbar^{2}$ and $(\ubar^{2}-\vbar^{2})^{2}$.
Owing to this symmetry, the primed kernel functions of the two members of each
pair of contractions (rows $\{1,2\}$, $\{3,4\}$ and $\{5,6\}$ of
table~\ref{tab:wick}) coincide exactly, even for a general spectrum.
Grouping the six terms into three pairs,
\begin{equation}\label{eq:IMDa}
	\begin{aligned}
		I^{(3),MD,1}_{j}&=2\,I^{(3),MD}_{j}(u,v,\ubar,\vbar,x)\,,\\
		I^{(3),MD,2}_{j}&=2\,I^{(3),MD}_{j}
		(\vbar v,w,u/w,\ubar v/w,x)\,,\\
		I^{(3),MD,3}_{j}&=2\,I^{(3),MD}_{j}
		(\ubar v,\sigma,u/\sigma,\vbar v/\sigma,x)\,,
	\end{aligned}
\end{equation}
with $w\equiv|\kb-\qb|/k$ and
$\sigma\equiv\sqrt{1+u^{2}-w^{2}+v^{2}(\ubar^{2}+\vbar^{2}-1)}$,
and denoting the pair sums of the momentum polynomials by
\begin{equation}\label{eq:Paij}
	\Pb^{ij}_{1}=\Pb^{ij}_{(1)}+\Pb^{ij}_{(2)}\,,\qquad
	\Pb^{ij}_{2}=\Pb^{ij}_{(3)}+\Pb^{ij}_{(4)}\,,\qquad
	\Pb^{ij}_{3}=\Pb^{ij}_{(5)}+\Pb^{ij}_{(6)}\,.
\end{equation}
Applying Wick's theorem and using the resulting two three-dimensional delta functions to eliminate the integrations over $\pb'$ and $\qb'$, we obtain
\begin{align}\label{eq:P6d}
	\Psf^{ij}_{h}(\eta,k)&=\frac{k^{3}}{32\pi^{2}}
	\int\frac{\dd^{3}p\,\dd^{3}q}
	{|\kb-\pb|^{3}|\pb-\qb|^{3}q^{3}}
	P_{\zeta}(ku)P_{\zeta}(\ubar vk)P_{\zeta}(\vbar vk)\nonumber\\
	&\times I^{(3),MD}_{i}(u,v,\ubar,\vbar,x)
	\sum_{a=1}^{3}\Pb^{ij}_{a}\,I^{(3),MD,a}_{j}\,.
\end{align}

\subsection{Reduction of the integration measure}\label{sec:measure}

We now reduce the six-dimensional phase space to a five-fold integral.
Choose $\kb$ along the $z$-axis, $\pb$ in the $x$-$z$ plane, and let
$\Delta\varphi$ be the azimuthal angle of $\qb$ around $\kb$.
The angular integrals are traded for $(u,w,\ubar)$ via
\begin{equation}\label{eq:angles}
	\cos\theta_{p}=\frac{1+v^{2}-u^{2}}{2v}\,,\qquad
	\cos\theta_{q}=\frac{1+\vbar^{2}v^{2}-w^{2}}{2\vbar v}\,,\qquad
	\hat\pb\cdot\hat\qb=\frac{1+\vbar^{2}-\ubar^{2}}{2\vbar}\,,
\end{equation}
and the six-dimensional measure becomes the five-fold form
\begin{equation}\label{eq:measure}
	\dd^{3}p\,\dd^{3}q
	=\frac{16\pi k^{6}v^{5}\vbar\,uw\ubar}
	{\sqrt{Y(1-X^{2})}}\,
	\dd v\,\dd\vbar\,\dd w\,\dd u\,\dd\ubar\,,
\end{equation}
where the $\pb$ azimuth gives $2\pi$, the two branches of
$\Delta\varphi$ give a factor $2$, and
\begin{equation}\label{eq:XY}
	X=\cos\Delta\varphi
	=\frac{\hat\pb\cdot\hat\qb-\cos\theta_{p}\cos\theta_{q}}
	{\sin\theta_{p}\sin\theta_{q}}\,,\qquad
	Y=16\,v^{4}\vbar^{2}\sin^{2}\theta_{p}\sin^{2}\theta_{q}\,.
\end{equation}
The integration domain is
\begin{align}\label{eq:domain}
	&v\in[0,\infty)\,,\quad \vbar\in[0,\infty)\,,\quad
	u\in\big[|1-v|,\,1+v\big]\,,\nonumber\\
    \nonumber\\
	&w\in\big[|1-\vbar v|,\,1+\vbar v\big]\,,\quad
	\ubar\in[\ubar_{-},\ubar_{+}]\,,
\end{align}
with
\begin{align}\label{eq:ubarpm}
	\ubar_{\pm}&=\left(1+\vbar^{2}
	-\frac{(1+v^{2}-u^{2})(1+\vbar^{2}v^{2}-w^{2})}{2v^{2}} \right. \nonumber\\
	&\left.\pm2\vbar\sqrt{\left[1-\left(\frac{1+v^{2}-u^{2}}{2v}\right)^{2}\right]
	\left[1-\left(\frac{1+\vbar^{2}v^{2}-w^{2}}{2\vbar v}\right)^{2}\right]}
	\,\right)^{1/2}.
\end{align}
We checked the measure \eqref{eq:measure} both analytically
($\int\dd^{3}p\,\dd^{3}q\,e^{-p^{2}-q^{2}}=\pi^{3}$ with $k=1$, and the
$\ubar$ integral of the measure alone gives $\pi\vbar$) and by an
independent Monte Carlo integration.

\subsection{Master formula}\label{sec:masterf}

Inserting Eq.~\eqref{eq:measure} into Eq.~\eqref{eq:P6d}, the power spectrum of third-order SIGW during the MD era for a general primordial spectrum is given by
\begin{equation}\label{eq:master}
	\begin{aligned}
		\Psf^{ij}_{h}(\eta,k)&=\frac{1}{2\pi}
		\int_{0}^{\infty}\dd v\int_{0}^{\infty}\dd\vbar
		\int_{|1-\vbar v|}^{1+\vbar v}\dd w
		\int_{|1-v|}^{1+v}\dd u\int_{\ubar_{-}}^{\ubar_{+}}\dd\ubar\;
		\frac{w}{u^{2}\ubar^{2}v\vbar^{2}\sqrt{Y(1-X^{2})}}\\
		&\times P_{\zeta}(ku)P_{\zeta}(\ubar vk)P_{\zeta}(\vbar vk) I^{(3),MD}_{i}(u,v,\ubar,\vbar,x)
		\sum_{a=1}^{3}\Pb^{ij}_{a}\,I^{(3),MD,a}_{j}\,,
	\end{aligned}
\end{equation}
where the indices $i$ and $j$ label the different source terms contributing to the third-order SIGWs, while the summation over the index $a$ arises from the different variable substitutions generated by the Wick expansion. The third-order kernel functions $I^{(3),MD,a}_{j}$ are given in Eq.~\eqref{eq:IMDa} and Eq.~\eqref{eq:I1}--Eq.~\eqref{eq:I4}. The momentum polynomials $\Pb^{ij}_{a}$ are computed from \eqref{eq:Pijdef}. More precisely, all scalar products of the five vectors $\{\hat\kb,\pb,\qb,\pb',\qb'\}$ and all contractions involving the transverse-traceless operator $\Lam_{lm,rs}$ can be reduced, using  table~\ref{tab:wick} and Eq.~\eqref{eq:angles}, to polynomials in $(u,v,w,\ubar,\vbar)$, so every $\Pb^{ij}_{a}$ is a rational polynomial
in these variables times $k^{4}$. Some useful explicit expressions and
properties are collected in appendix~\ref{Ap:poly}.
The master formula \eqref{eq:master} is the MD counterpart of the
radiation-domination formula of Ref.~\cite{Zhou:2021vcw}: the
integration measure, the domain, and the momentum polynomials are
unchanged, and the only difference is the replacement of the kernel
functions by their MD counterparts.
The total power spectrum of the third-order SIGW is $\Psf^{(3)}_{h}=\sum_{i,j=1}^{4}\Psf^{ij}_{h}$.

\paragraph{Exact symmetry of the cross power spectra.} Since relabelling $(\pb,\qb,\kb)\leftrightarrow(\pb',\qb',\kb')$ is a
dummy change of integration variables in Eq.~\eqref{eq:Ppij}, the cross
spectra obey the exact symmetry
\begin{equation}\label{eq:Psym}
	\Psf^{ij}_{h}(\eta,k)=\Psf^{ji}_{h}(\eta,k)\,,\qquad
	\Psf^{(3)}_{h}=\sum_{i=1}^{4}\Psf^{ii}_{h}
	+2\sum_{i<j}\Psf^{ij}_{h}\,.
\end{equation}
Specifically, the equality is most transparent when $\Psf^{ij}_{h}$ is evaluated by
eliminating $(\pb',\qb')$ with the three-dimensional $\delta$ functions while
$\Psf^{ji}_{h}$ is evaluated by eliminating $(\pb,\qb)$ instead: the
two reduced expressions are then formally identical. If both are
reduced keeping $\dd^{3}p\,\dd^{3}q$, the integrands differ pointwise
($\Pb^{ij}_{a}\neq\Pb^{ji}_{a}$ for $a=2,3$) and Eq.~\eqref{eq:Psym}
holds only after integration. We have verified Eq.~\eqref{eq:Psym}
numerically at the level of the full five-fold integral, as detailed in
section~\ref{sec:mono}. This symmetry is not only a nontrivial
self-consistency check of the master formula, but also halves the computational cost.

\subsection{Energy density spectrum}\label{sec:omega}

With the perturbative expansion $h_{ij}=\sum_{n}\frac{1}{n!}
h^{(n)}_{ij}$, the physical spectrum is
\begin{equation}\label{eq:Phphys}
	\Psf^{\mathrm{tot}}_{h}(\eta,k)=\frac14\,\Psf^{(2)}_{h}(\eta,k)
	+\frac{1}{36}\,\Psf^{(3)}_{h}(\eta,k)
	+O\!\left(\Psf^{(4)}_{h}\right),
\end{equation}
where the factorial coefficients originate from the metric expansion
\eqref{eq:metric}. The fractional energy density of gravitational waves is
\begin{equation}\label{eq:Omega}
	\Omega_{\mathrm{GW}}(\eta,k)
	=\frac{1}{6}\left(\frac{k}{\mathcal{H}}\right)^{2}
	\overline{\Psf^{\mathrm{tot}}_{h}(\eta,k)}\,,
\end{equation}
where the overline denotes the time average over the oscillation
period. During matter domination, however, the induced tensor modes do
not oscillate but grow secularly; as emphasized in
Ref.~\cite{Kohri:2018awv}, the oscillation-average prescription then
overestimates the time-averaged energy density by a factor of two,
which we compensate by an extra factor $1/2$. With $\mathcal{H}=2/\eta$, we arrive at
\begin{equation}\label{eq:OmegaMD}
	\Omega^{(2)}_{\mathrm{GW}}(\eta,k)
	=\frac{x^{2}}{192}\,\overline{\Psf^{(2)}_{h}(\eta,k)}\,,\qquad
	\Omega^{(3)}_{\mathrm{GW}}(\eta,k)
	=\frac{x^{2}}{1728}\,\overline{\Psf^{(3)}_{h}(\eta,k)}\,.
\end{equation}
The above results provide the energy density spectra of the second-order and third-order SIGWs during the MD era.

\section{Monochromatic primordial power spectrum}\label{sec:mono}

In this section, we evaluate the master formula \eqref{eq:master} for
the simplest benchmark, a monochromatic primordial spectrum. We reduce
the master formula to a two-fold integral in
section~\ref{sec:monored}, prove in section~\ref{sec:h3zero} that all
contributions involving the second-order vector source vanish
identically on the monochromatic support, and present the numerical
results in section~\ref{sec:mononum}.

\subsection{Reduction of the master formula}\label{sec:monored}

For the monochromatic primordial spectrum
\begin{equation}\label{eq:mono}
	P_{\zeta}(k)=\mathcal{A}_{\zeta}\,k_{*}\,\delta(k-k_{*})\,,
\end{equation}
the three $\delta$ functions in Eq.~\eqref{eq:master} fix
\begin{equation}\label{eq:monosub}
	u=\frac{1}{\tk}\,,\qquad
	\ubar=\vbar=\frac{1}{v\tk}\,,\qquad
	\tk\equiv\frac{k}{k_{*}}\,,
\end{equation}
where $\tk=k/k_*$ and the five-fold integral \eqref{eq:master} reduces to the two-fold form
\begin{equation}\label{eq:master-mono}
	\begin{aligned}
		\Psf^{ij}_{h}(x,\tk)=&\frac{\mathcal{A}_{\zeta}^{3}\,\tk^{3}}{2\pi}\,
		\Theta(3-\tk)
		\int_{\left|1-\tk^{-1}\right|}^{\min\{2\tk^{-1},\,1+\tk^{-1}\}}
		\dd v\int_{w_{-}}^{w_{+}}\dd w\;
		&\frac{vw}{\sqrt{Y(1-X^{2})}}\,
		I^{(3),MD}_{i}\Big|_{\mathrm{mono}}\\
		&\times\sum_{a=1}^{3}\Pb^{ij}_{a}\,
		I^{(3),MD,a}_{j}\Big|_{\mathrm{mono}}\,,
	\end{aligned}
\end{equation}
where the kernel functions are evaluated with Eq.~\eqref{eq:monosub} and
\begin{equation}\label{eq:IMDa-mono}
	\begin{aligned}
		I^{(3),MD,1}_{j}&=2\,I^{(3),MD}_{j}\left(\frac{1}{\tk},v,
		\frac{1}{v\tk},\frac{1}{v\tk},x\right),\\
		I^{(3),MD,2}_{j}&=2\,I^{(3),MD}_{j}\left(\frac{1}{\tk},w,
		\frac{1}{w\tk},\frac{1}{w\tk},x\right),\\
		I^{(3),MD,3}_{j}&=2\,I^{(3),MD}_{j}\left(\frac{1}{\tk},
		\sigma_{m},\frac{1}{\tk\sigma_{m}},\frac{1}{\tk\sigma_{m}},
		x\right),\qquad
		\sigma_{m}=\sqrt{1-v^{2}-w^{2}+\frac{3}{\tk^{2}}}\,,
	\end{aligned}
\end{equation}
and the remaining integration range is
\begin{equation}\label{eq:wpm}
	w_{\pm}=\left(\frac{1}{2}+\frac{3}{2\tk^{2}}-\frac{v^{2}}{2}
	\pm\frac{1}{2v}\sqrt{\left(v^{2}-\frac{4}{\tk^{2}}\right)
	\left(v^{2}-\left(1-\frac{1}{\tk}\right)^{2}\right)
	\left(v^{2}-\left(1+\frac{1}{\tk}\right)^{2}\right)}\,
	\right)^{1/2}.
\end{equation}
The upper cutoff $\tk<3$ reflects the kinematic support of the monochromatic convolution: three wavevectors of magnitude $k_*$ can sum to $\mathbf{k}$ only if $k<3k_*$, whereas the corresponding bound is $k<2k_*$ at second order.

\subsection{Vanishing of the vector-induced contributions}
\label{sec:h3zero}

On the monochromatic support \eqref{eq:monosub}, the vector-induced
momentum structure summed over each Wick pair sharing the same kernel function vanishes
identically,
\begin{equation}\label{eq:S3id}
	\begin{aligned}
		\Ssf_{3}(-\pb,-\qb)+\Ssf_{3}(-\pb,\qb-\pb)&=0\,,\\
		\Ssf_{3}(\qb-\kb,\qb-\pb)+\Ssf_{3}(\qb-\kb,\pb-\kb)&=0\,,\\
		\Ssf_{3}(\pb-\qb-\kb,-\qb)+\Ssf_{3}(\pb-\qb-\kb,\pb-\kb)&=0\,,
	\end{aligned}
\end{equation}
i.e.\ $\Pb^{i3}_{a}=0$ pointwise for all $i=1,\dots,4$ and $a=1,2,3$.
Each identity follows from $P_{lm}(\hat\pb)p_{m}=0$ together with the
equal-magnitude conditions $q=|\pb-\qb|=|\kb-\pb|=k_{*}$; for instance,
the first pair is proportional to
$2\,\pb\cdot\qb-p^{2}=q^{2}-|\pb-\qb|^{2}=0$. The details are given in
appendix~\ref{Ap:poly}. Consequently, for a monochromatic primordial
spectrum
\begin{equation}\label{eq:h3zero}
	\Psf^{i3}_{h}=\Psf^{3i}_{h}=0\qquad(i=1,\dots,4)\,,
\end{equation}
where the second equality follows from Eq.~\eqref{eq:Psym}. Hence
$h^{(3)}_{ij,3}$, the contribution induced by the second-order vector
modes, drops out of the monochromatic third-order spectrum entirely,
including $\Psf^{33}_{h}$, and
\begin{equation}\label{eq:Pmono-red}
	\Psf^{(3)}_{h}\Big|_{\mathrm{mono}}
	=\sum_{i,j\in\{1,2,4\}}\Psf^{ij}_{h}
	=\sum_{i\in\{1,2,4\}}\Psf^{ii}_{h}
	+2\,\big(\Psf^{12}_{h}+\Psf^{14}_{h}+\Psf^{24}_{h}\big)\,.
\end{equation}
We emphasize that Eq.~\eqref{eq:S3id} is special to the monochromatic
support; for a general spectrum all sixteen terms contribute, as will
be the case in section~\ref{sec:pbh}.

This result also provides a stringent test of our numerics. The
manifestly vanishing combinations \eqref{eq:S3id} are obtained by
eliminating the primed momenta $(\pb',\qb')$ with the $\delta$
functions; if one instead evaluates $\Psf^{3i}_{h}$ keeping
$\dd^{3}p'\,\dd^{3}q'$, the integrand is nonzero pointwise (of the same
order as the dominant terms) and integrates to zero through a delicate
cancellation. Numerically we find that the pairs $(i,3)$ vanish
pointwise at the integrand level (of order $10^{-9}$ compared with the
$10^{6}$ magnitude of the dominant integrands), while the pairs $(3,i)$
integrate to $10^{-7}$ in absolute terms, i.e.\ $10^{-13}$ relative to
the total, and the residual does not grow with resolution, consistent
with an exact cancellation. Likewise, the symmetry \eqref{eq:Psym} is
verified explicitly: $\Psf^{12}_{h}$ and $\Psf^{21}_{h}$ agree to a
relative accuracy of $2\times10^{-6}$, while $\Psf^{14}_{h}$ versus
$\Psf^{41}_{h}$ and $\Psf^{24}_{h}$ versus $\Psf^{42}_{h}$ agree to
$5\times10^{-5}$ and $3\times10^{-4}$, respectively, with the residuals
shrinking as the grid resolution increases.

\subsection{Numerical results}\label{sec:mononum}

\begin{figure}[t]
	\centering
	\includegraphics[width=\textwidth]{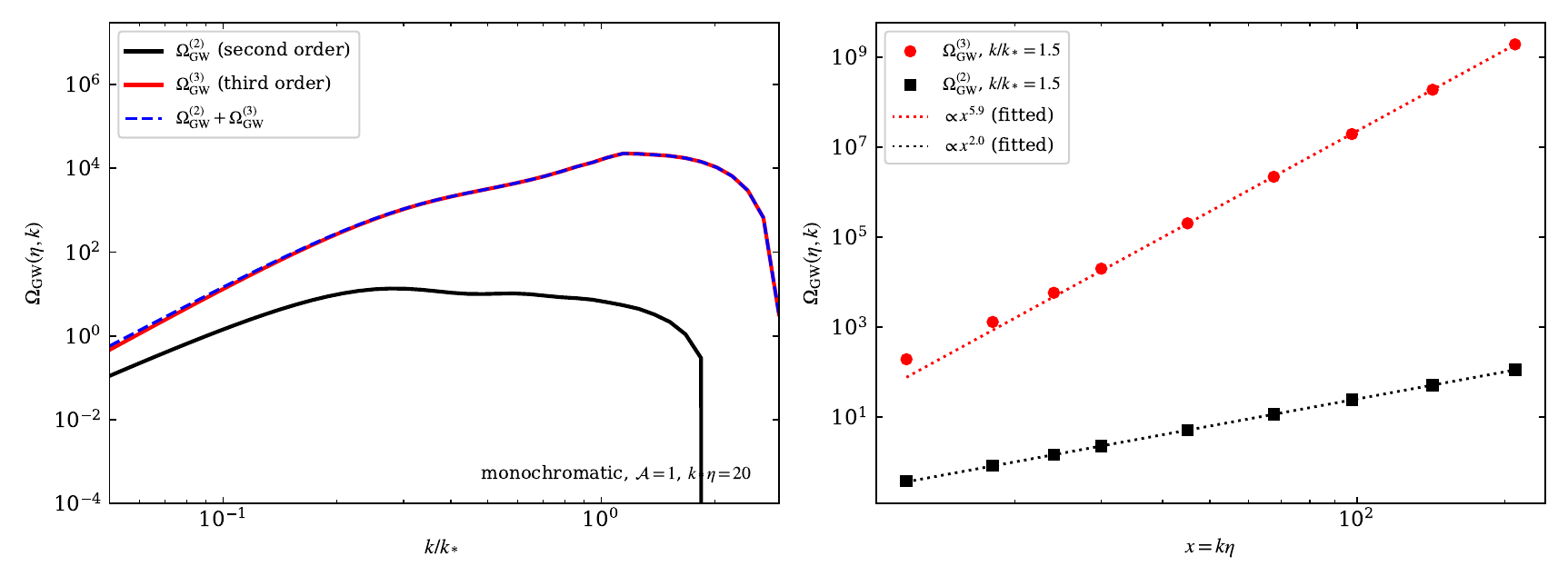}
	\caption{\emph{Left panel}: energy density spectra of the second-order
	and third-order SIGWs during matter domination for the
	monochromatic primordial spectrum \eqref{eq:mono} with
	$\mathcal{A}_{\zeta}=1$, evaluated at $k_{*}\eta=20$. The third-order
	spectrum is supported on $k<3k_{*}$, peaks at $k\simeq1.14\,k_{*}$,
	and dominates the second-order one (supported on $k<2k_{*}$) over
	most of the range already at this moderate time.  \emph{Right panel}: time dependence at the representative
	wavenumber $k=1.5\,k_{*}$. The dotted lines are power-law fits to
	the late-time points, giving $\Omega^{(3)}_{\mathrm{GW}}\propto
	x^{5.94}$ and $\Omega^{(2)}_{\mathrm{GW}}\propto x^{2.00}$, in
	agreement with the analytic expectations $x^{6}$ and $x^{2}$.}
	\label{fig:mono}
\end{figure}

We evaluate Eq.~\eqref{eq:master-mono} numerically with the Gaussian
value $a_{\mathrm{NL}}=1$. The details of the numerical
implementation are given in appendix~\ref{Ap:num}. As a benchmark, our
second-order pipeline agrees with the asymptotic analytic MD result of
Ref.~\cite{Kohri:2018awv} in its domain of validity $x\gg1$,
\begin{equation}\label{eq:KT40}
	\Omega^{(2)}_{\mathrm{GW}}(\eta,k)
	=\frac{3}{25}\left(\frac{k_{*}}{aH}\right)^{2}
	\left(1-\frac{k^{2}}{4k_{*}^{2}}\right)^{2}
	\mathcal{A}_{\zeta}^{2}\,\Theta(2k_{*}-k)\,.
\end{equation}
The results are shown in figure~\ref{fig:mono}. At $k_{*}\eta=20$ the
third-order spectrum peaks at $k\simeq1.14\,k_{*}$ with
$\Omega^{(3)}_{\mathrm{GW}}\simeq2.3\times10^{4}\,\mathcal{A}_{\zeta}^{3}$,
while the second-order spectrum peaks at $13.6\,\mathcal{A}_{\zeta}^{2}$; the
third-order contribution already dominates over most of the range and
extends to $k=3k_{*}$. The dominant contribution comes from the fourth
class of sources, i.e.\ the product of the first-order scalar with the
second-order scalars, whose diagonal term $\Psf^{44}_{h}$ alone
saturates the peak; the cross terms $\Psf^{14}_{h}$ and $\Psf^{24}_{h}$
are negative and positive respectively and partially cancel.

Furthermore, the right panel of
figure~\ref{fig:mono} shows that the third-order spectrum grows as
$\Omega^{(3)}_{\mathrm{GW}}\propto x^{5.94}$ (fitted), in agreement
with the analytic expectation $x^{6}$: the kernel functions grow as
$I^{(3)}\propto x^{2}$, hence
$\Psf^{(3)}_{h}\propto x^{4}$ and $\Omega^{(3)}_{\mathrm{GW}}\propto
x^{2}\Psf^{(3)}_{h}\propto x^{6}$. This should be contrasted with the
second-order result $\Omega^{(2)}_{\mathrm{GW}}\propto x^{2}$, whose
kernel function \eqref{eq:Ih2} asymptotes to a constant. The ratio therefore grows as
\begin{equation}\label{eq:Rmono}
	\frac{\Omega^{(3)}_{\mathrm{GW}}}{\Omega^{(2)}_{\mathrm{GW}}}
	\propto \mathcal{A}_{\zeta}\,x^{4}\,,
\end{equation}
so that the perturbative hierarchy is inevitably inverted during a
sufficiently long MD era, no matter how small the primordial amplitude
is. 

\section{Application to a PBH-dominated era}\label{sec:pbh}

As an application of the master formula, we now turn to the early MD
era driven by ultralight primordial black holes. We briefly review the
model and the second-order backreaction bound in
section~\ref{sec:pbhmodel}, compute the third-order SIGW spectrum at
evaporation in section~\ref{sec:pbh3}, and derive the resulting bound
on the initial PBH abundance in section~\ref{sec:pbhbound}.

\subsection{PBH-dominated era}\label{sec:pbhmodel}

We apply our results to the early MD era induced by a domination of
ultralight PBHs, following the model of Ref.~\cite{Papanikolaou:2020qtd}
(see also Refs.~\cite{Domenech:2020ssp,Papanikolaou:2022chm,Inomata:2020lmk}). Ultralight PBHs of mass $m_{\mathrm{PBH}}$ are formed from enhanced
small-scale curvature perturbations with an initial abundance
$\Omega_{\mathrm{PBH},f}$, the fraction of the total energy density in
PBHs at formation. The PBH mass is related to the conformal Hubble
scale at formation by $m_{\mathrm{PBH}}=4\pi M_{\mathrm{Pl}}^{2}/H_{f}$
(up to the formation-efficiency factor, which we absorb into the
definition of $m_{\mathrm{PBH}}$). Since PBHs redshift as matter while
the radiation background redshifts faster, they dominate the universe at
the Hubble scale
\begin{equation}\label{eq:Hd}
	H_{d}=\Omega_{\mathrm{PBH},f}\,H_{f}\,,
\end{equation}
provided that domination occurs before evaporation,
\begin{equation}\label{eq:dom}
	\Omega_{\mathrm{PBH},f}>10^{-15}
	\left(\frac{10^{9}\,\mathrm{g}}{m_{\mathrm{PBH}}}\right),
\end{equation}
which is eq.~(4.2) of Ref.~\cite{Papanikolaou:2020qtd}. The PBHs then
evaporate through Hawking radiation, and the duration of the
PBH-dominated era in units of the Hubble time at domination is
\begin{equation}\label{eq:tau}
	\tau\equiv\frac{t_{\mathrm{evap}}}{t_{d}}
	=\frac{1280}{g_{*}}\left(\frac{m_{\mathrm{PBH}}}{M_{\mathrm{Pl}}}\right)^{2}
	\Omega_{\mathrm{PBH},f}^{2}\,,
\end{equation}
with $g_{*}$ the effective number of relativistic degrees of freedom,
taken to be $100$, and $M_{\mathrm{Pl}}=4.342\times10^{-6}\,\mathrm{g}$
the reduced Planck mass. The model is therefore fully specified by the
two parameters $(m_{\mathrm{PBH}},\Omega_{\mathrm{PBH},f})$. During the PBH-dominated era, the power spectrum of the Bardeen
potential is given by \cite{Papanikolaou:2020qtd}
\begin{equation}\label{eq:PPhiPBH}
	P_{\Phi}(k)=\frac{2}{3\pi}\left(\frac{k}{k_{\mathrm{UV}}}\right)^{3}
	\left[5+\frac{4k^{2}}{9H_{d}^{2}}\right]^{-2},
\end{equation}
where the ultraviolet cutoff scale
\begin{equation}\label{eq:kUV}
	k_{\mathrm{UV}}=H_{f}\,\Omega_{\mathrm{PBH},f}^{1/3}
	=H_{d}\,\Omega_{\mathrm{PBH},f}^{-2/3}
\end{equation}
corresponds to the horizon scale at formation. All comoving wavenumbers
are normalized to the comoving Hubble scale at domination, and we
define
\begin{equation}\label{eq:ydef}
	y_d\equiv\frac{k}{H_{d}}\,,\qquad
	\frac{k}{a_{\mathrm{evap}}H_{\mathrm{evap}}}=y_d\,\tau^{1/3}\,,\qquad
	x=k\eta_{\mathrm{evap}}=2y_d\,\tau^{1/3}\,,
\end{equation}
where the last two relations follow from the MD background evolution
between domination and evaporation. The spectrum \eqref{eq:PPhiPBH}
peaks around $k\sim k_{\mathrm{UV}}\sim\Omega_{\mathrm{PBH},f}^{-2/3}
H_{d}$ with an amplitude $\sim\Omega_{\mathrm{PBH},f}^{2}$. When
evaluating the master formula \eqref{eq:master}, which is written in
terms of the primordial curvature perturbation, we use
$P_{\zeta}(k)=(25/9)P_{\Phi}(k)$ following Eq.~\eqref{eq:phi1}.

The second-order SIGWs produced during the PBH-dominated era grow
secularly and reach their maximal value at evaporation. Requiring that
they do not backreact on the cosmological perturbations,
$\int\dd\ln k\;\Omega^{(2)}_{\mathrm{GW}}
	(\eta_{\mathrm{evap}},k)<1$, Ref.~\cite{Papanikolaou:2020qtd}
derived the upper bound
\begin{equation}\label{eq:bound2}
	\Omega_{\mathrm{PBH},f}\lesssim1.4\times10^{-4}
	\left(\frac{10^{9}\,\mathrm{g}}{m_{\mathrm{PBH}}}\right)^{1/4}\ .
\end{equation}
Combined with the domination condition
\eqref{eq:dom}, this leaves a wide window of PBH-dominated cosmologies. In the subsequent discussion, we consider the corrections to this constraint contributed by third-order SIGWs.

\subsection{The third-order correction}\label{sec:pbh3}

\begin{figure}[t]
	\centering
	\includegraphics[width=0.75\textwidth]{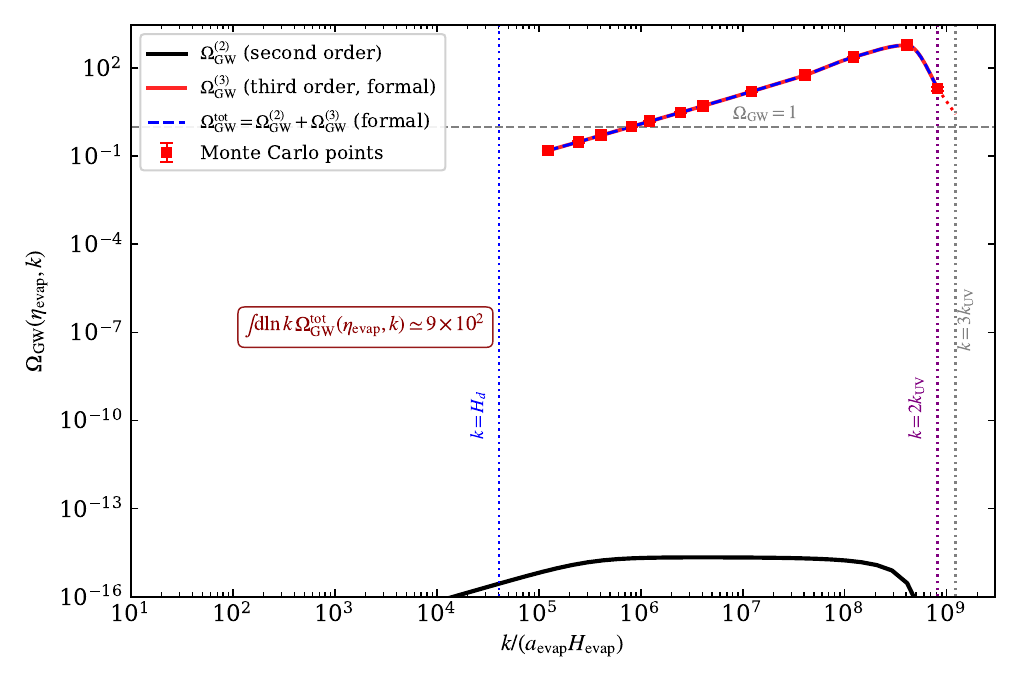}
	\caption{Energy density spectra of the second-order and third-order
	SIGWs at PBH evaporation for $m_{\mathrm{PBH}}=10^{7}\,\mathrm{g}$
	and $\Omega_{\mathrm{PBH},f}=10^{-6}$. The black curve is the
	second-order result, which plateaus at $\sim2\times10^{-15}$. The
	red squares with error bars are our Monte Carlo evaluations of the
	full third-order expression \eqref{eq:master}, the red curve is a
	smooth interpolation, and the red dotted segment extrapolates the
	measured power-law tail up to the edge of the third-order support
	$k=3k_{\mathrm{UV}}$; the blue dashed curve is the formal sum of
	the two contributions. The vertical dotted lines mark the
	domination scale $k=H_{d}$ and the cutoff scales
	$k=2k_{\mathrm{UV}}$ and $k=3k_{\mathrm{UV}}$, and the horizontal
	dashed line marks $\Omega_{\mathrm{GW}}=1$. The boxed value is the
	logarithmic integral of the total spectrum at evaporation, which
	formally exceeds unity by almost three orders of magnitude; the
	third-order curve must therefore be understood as a \emph{formal}
	perturbative result signaling the breakdown of the perturbative
	expansion for these parameters, see the discussion in the text.}
	\label{fig:pbhspec}
\end{figure}

We now compute the third-order energy density spectrum at evaporation by inserting the
spectrum \eqref{eq:PPhiPBH} into the master formula \eqref{eq:master}.
Since the spectrum is broad, spanning four decades from $H_{d}$ to
$k_{\mathrm{UV}}$, all sixteen terms $\Psf^{ij}_{h}$ contribute, and the
five-fold integral with its inverse-square-root measure singularities is
evaluated by a six-dimensional Cartesian Monte Carlo integration,
cross-checked against the five-fold grid at intermediate points (see
appendix~\ref{Ap:num}).

The result for the benchmark parameters $m_{\mathrm{PBH}}=10^{7}\,
\mathrm{g}$, $\Omega_{\mathrm{PBH},f}=10^{-6}$ (for which
$\tau^{1/3}=4.08\times10^{4}$) is shown in figure~\ref{fig:pbhspec}.
The second-order spectrum plateaus at $\sim2\times10^{-15}$, in
agreement with Ref.~\cite{Papanikolaou:2020qtd}. The third-order
spectrum, in contrast, grows monotonically across the support and
reaches $\Omega^{(3)}_{\mathrm{GW}}\sim0.5$ at $y_d=10$ and $\sim57$ at
$y_d=10^{3}$, peaks around the cutoff scale $y_d\simeq
k_{\mathrm{UV}}/H_{d}=10^{4}$ at $\Omega^{(3)}_{\mathrm{GW}}
\simeq6\times10^{2}$, and drops steeply beyond the peak, approximately
as $y_d^{-4.9}$ over the last sampled interval. While the second-order
spectrum vanishes beyond $k=2k_{\mathrm{UV}}$, the third-order one is
supported up to $k=3k_{\mathrm{UV}}$. The logarithmic integral of the
total spectrum at evaporation is then dominated entirely by the
third-order contribution and formally evaluates to
\begin{equation}\label{eq:Itot}
	\int\dd\ln k\;\Omega^{\mathrm{tot}}_{\mathrm{GW}}
	(\eta_{\mathrm{evap}},k)\simeq9\times10^{2}\,,
\end{equation}
to be compared with $1.5\times10^{-14}$ from the second-order spectrum
alone. The huge enhancement over the second-order plateau is a direct
consequence of the different time dependence of the two spectra: while
the second-order spectrum grows as $\Omega^{(2)}_{\mathrm{GW}}\propto
x^{2}$ and its kernel function \eqref{eq:Ih2} asymptotes to a constant, the third-order
spectrum grows as $\Omega^{(3)}_{\mathrm{GW}}\propto x^{6}$ because of
the secular kernel functions \eqref{eq:Ilate}, and the evaporation time
corresponds to $x\sim10^{9}$ at the peak scale.

Since the total induced energy density \eqref{eq:Itot} far exceeds
unity for these parameters, the perturbative expansion has broken
down, and the third-order curve in figure~\ref{fig:pbhspec} must be
understood as a formal extrapolation of the perturbative formulas. The
backreaction requirement applied to the total spectrum then yields a
much stronger bound on the initial PBH abundance than the second-order
analysis, as we now quantify.

\subsection{Backreaction bound on the PBH abundance}\label{sec:pbhbound}

\begin{figure}[t]
	\centering
	\includegraphics[width=0.75\textwidth]{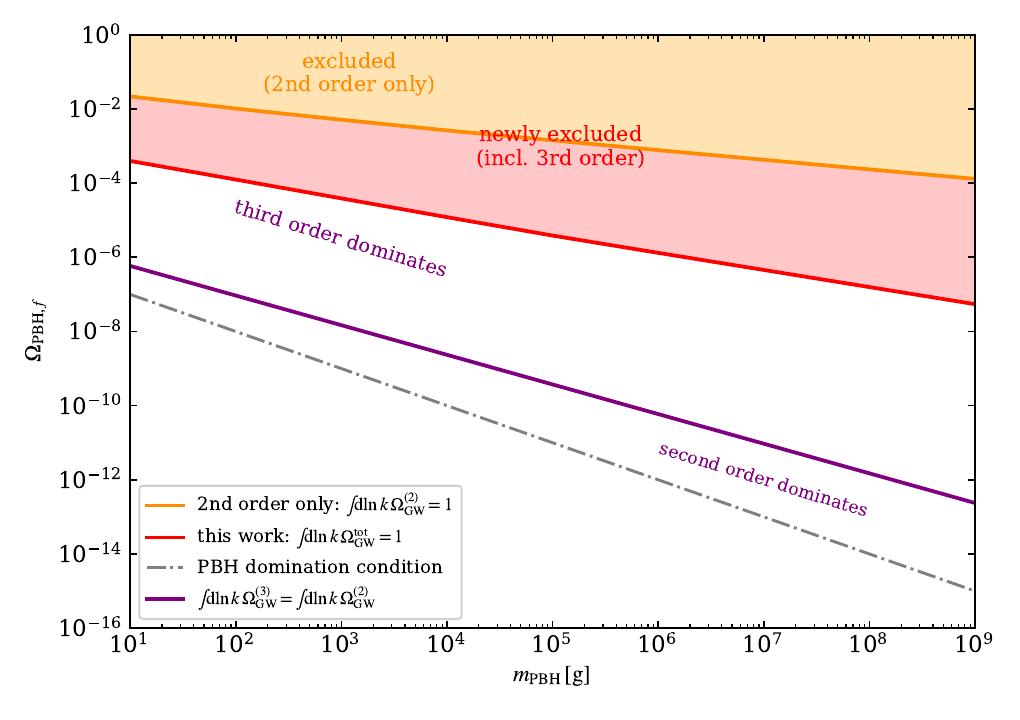}
	\caption{Constraints on the initial PBH abundance
	$\Omega_{\mathrm{PBH},f}$ as a function of the PBH mass. The orange
	shaded region is excluded by the second-order-only backreaction
	criterion $\int\mathrm{d}\ln k\,\Omega^{(2)}_{\mathrm{GW}}
	(\eta_{\mathrm{evap}},k)=1$ of Ref.~\cite{Papanikolaou:2020qtd},
	which we reproduce numerically. The red line is the boundary
	\eqref{eq:bound3} obtained in this work from the same criterion
	applied to the total spectrum,
	$\int\mathrm{d}\ln k\,\Omega^{\mathrm{tot}}_{\mathrm{GW}}
	(\eta_{\mathrm{evap}},k)=1$ with $\Omega^{\mathrm{tot}}_{\mathrm{GW}}
	=\Omega^{(2)}_{\mathrm{GW}}+\Omega^{(3)}_{\mathrm{GW}}$; the red
	shaded region between the two boundaries is newly excluded once the
	third-order contribution is included. The gray dash-dotted line is
	the PBH domination condition \eqref{eq:dom}. The purple line marks where the integrated third-order and second-order
energy densities are equal in Eq.~\eqref{eq:hier}; between it and the
red line the third order dominates while the total energy density of SIGW is still small compared with the background.}
	\label{fig:pbhbound}
\end{figure}

We now derive the bound on the initial PBH abundance. We adopt the
same backreaction criterion as the second-order analysis of
Ref.~\cite{Papanikolaou:2020qtd}, namely that the total SIGW energy
density at evaporation does not exceed the total energy density of the
universe,
\begin{equation}\label{eq:crit}
	\int\dd\ln k\;\Omega^{\mathrm{tot}}_{\mathrm{GW}}
	(\eta_{\mathrm{evap}},k)<1\,,\qquad
	\Omega^{\mathrm{tot}}_{\mathrm{GW}}=\Omega^{(2)}_{\mathrm{GW}}
	+\Omega^{(3)}_{\mathrm{GW}}\,.
\end{equation}
At the benchmark mass $m_{\mathrm{PBH}}=10^{7}\,\mathrm{g}$ and the
reference abundance $\Omega_{\mathrm{PBH},f}=10^{-6}$, the integral in
Eq.~\eqref{eq:crit} evaluates to $9\times10^{2}$, far above unity;
solving for the abundance at which it equals one gives the critical
value $\Omega_{\mathrm{PBH},f}=4.6\times10^{-7}$, and repeating the
computation across the mass range gives
\begin{equation}\label{eq:bound3}
	\Omega_{\mathrm{PBH},f}\lesssim4.6\times10^{-7}
	\left(\frac{m_{\mathrm{PBH}}}{10^{7}\,\mathrm{g}}\right)^{-0.48},
\end{equation}
shown as the red line in figure~\ref{fig:pbhbound}. This strengthens
the second-order-only bound \eqref{eq:bound2} by a factor ranging from
$\sim50$ at $m_{\mathrm{PBH}}=10\,\mathrm{g}$ to $\sim2\times10^{3}$
at $m_{\mathrm{PBH}}=10^{9}\,\mathrm{g}$, i.e.\ about three orders of
magnitude at $m_{\mathrm{PBH}}=10^{7}\,\mathrm{g}$. Correspondingly,
the allowed window between the domination condition \eqref{eq:dom} and
the backreaction bound shrinks from $9.6$ to $6.7$ orders of magnitude
in $\Omega_{\mathrm{PBH},f}$ at $m_{\mathrm{PBH}}=10^{7}\,\mathrm{g}$.
The boundary \eqref{eq:bound3} intersects the domination condition
only at $m_{\mathrm{PBH}}\sim10^{-6}\,\mathrm{g}$, far below the mass
range of interest, so the window remains open for all
$m_{\mathrm{PBH}}\gtrsim10\,\mathrm{g}$.

It is also instructive to compare the integrated energy densities of
the two contributions directly. The logarithmic integrals scale as
$\int\mathrm{d}\ln k\,\Omega^{(3)}_{\mathrm{GW}}\propto
\Omega_{\mathrm{PBH},f}^{8.67}\,m_{\mathrm{PBH}}^{4}$ and
$\int\mathrm{d}\ln k\,\Omega^{(2)}_{\mathrm{GW}}\propto
\Omega_{\mathrm{PBH},f}^{16/3}\,m_{\mathrm{PBH}}^{4/3}$, normalized
to $9.0\times10^{2}$ and $1.5\times10^{-14}$, respectively, at the
benchmark point $m_{\mathrm{PBH}}=10^{7}\,$g,
$\Omega_{\mathrm{PBH},f}=10^{-6}$. They are equal along the purple
line in figure~\ref{fig:pbhbound},
\begin{equation}\label{eq:hier}
	\int\mathrm{d}\ln k\,\Omega^{(3)}_{\mathrm{GW}}
	=\int\mathrm{d}\ln k\,\Omega^{(2)}_{\mathrm{GW}}
	\quad\Longleftrightarrow\quad
	\Omega_{\mathrm{PBH},f}\simeq9.4\times10^{-12}
	\left(\frac{m_{\mathrm{PBH}}}{10^{7}\,\mathrm{g}}\right)^{-0.80}.
\end{equation}
Between the purple line and the red bound the integrated third-order
energy density exceeds the second-order one, while their sum is
still small compared with the background; below the line the
perturbative hierarchy is conventional. Since the line lies above
the domination condition \eqref{eq:dom} throughout the plotted
range, a PBH-dominated era with a conventional perturbative
hierarchy survives in the window between the gray and purple lines.

The boundary \eqref{eq:bound3} is
the point where the total induced energy density formally reaches
unity; since the third-order contribution alone is already of order
one there, still higher-order contributions are not controlled, and
the boundary should be understood as the point where the perturbative
SIGW calculation ceases to be predictive. This is the same logical
status as the second-order backreaction bound of
Ref.~\cite{Papanikolaou:2020qtd}, now reached at a much lower PBH abundance. Moreover, our evaluation samples the third-order spectrum up to
$y_d=2\times10^{4}$, where the last Monte Carlo point carries a $17\%$
statistical error, and extrapolates the remaining tail up to
$k=3k_{\mathrm{UV}}$ with the measured power law
$\Omega^{(3)}_{\mathrm{GW}}\propto y_d^{-4.9}$; we estimate an
$\mathcal{O}(10\%)$ systematic uncertainty on the normalization of
Eq.~\eqref{eq:bound3}, which does not affect our conclusions.

\section{Summary and discussion}\label{sec:summary}

We computed the third-order scalar induced gravitational waves during a MD era for the first time. More precisely, we derived the third-order equation of motion in Eq.~\eqref{eq:h3eom}
with the source classified into four parts in Eq.~\eqref{eq:Sdec}, built
respectively from three first-order scalars, and from the products of a
first-order scalar with the second-order tensor, vector, and scalar
perturbations. All four kernel functions were obtained analytically in
closed form in Eq.~\eqref{eq:I1}--Eq.~\eqref{eq:I4}, in terms of the two
elementary Green-function integrals \eqref{eq:J2} and \eqref{eq:M0}. Furthermore, the MD kernel functions exhibit a secular growth absent in radiation
domination: $\Jsf_{2}\propto x^{2}$ at late times, so three of the four
kernel functions grow as $I^{(3)}\propto x^{2}$. As a
consequence, the third-order energy density spectrum grows as
$\Omega^{(3)}_{\mathrm{GW}}\propto x^{6}$, whereas the second-order one
grows as $\Omega^{(2)}_{\mathrm{GW}}\propto x^{2}$; their ratio grows as
$\mathcal{A}_{\zeta}\,x^{4}$. The physical origin is transparent: during matter
domination the first-order scalar perturbations are frozen and the
induced tensor modes grow linearly with the scale factor, so each
additional perturbative order involving the growing modes brings in
additional powers of $x$. This is in sharp contrast to radiation
domination, where the sub-horizon scalar perturbations oscillate and
decay and the induced tensor modes oscillate with a constant amplitude.

We derived the master formula \eqref{eq:master} for the
third-order power spectrum as a five-fold integral for a general
primordial spectrum, proved the exact symmetry
$\Psf^{ij}_{h}=\Psf^{ji}_{h}$ of the cross spectra, and verified it
numerically. For the monochromatic primordial spectrum we showed that
all contributions involving the second-order vector source vanish
identically in Eq.~\eqref{eq:h3zero}, and evaluated the spectrum
numerically. The third-order spectrum is supported on $k<3k_{*}$,
peaks at $k\simeq1.14\,k_{*}$, and is dominated by the fourth class of sources $\mathcal{S}^{(3)}_{lm,4}$. Moreover, applying our results to the PBH-dominated early universe of
Ref.~\cite{Papanikolaou:2020qtd}, we found that the secular growth
makes the third-order contribution dominate the total SIGW energy
density at evaporation: at the reference abundance
$\Omega_{\mathrm{PBH},f}=10^{-6}$ for the benchmark mass
$m_{\mathrm{PBH}}=10^{7}\,\mathrm{g}$, the logarithmic integral of
the total spectrum evaluates to $9\times10^{2}$, far above unity.
The backreaction criterion $\int\mathrm{d}\ln k\,
\Omega^{\mathrm{tot}}_{\mathrm{GW}}(\eta_{\mathrm{evap}},k)<1$, i.e.\
the criterion of the second-order analysis now applied to the total
spectrum, then bounds the initial PBH abundance as
$\Omega_{\mathrm{PBH},f}\lesssim4.6\times10^{-7}\,
(m_{\mathrm{PBH}}/10^{7}\,\mathrm{g})^{-0.48}$, strengthening the
second-order-only bound by about three orders of magnitude and
shrinking the allowed window of PBH-dominated cosmologies
accordingly. The remaining window is
further structured by Eq.~\eqref{eq:hier}, where the integrated
third-order and second-order energy densities are equal: between it and
the bound the third order dominates while the total energy density
of SIGWs is still small compared with the background energy
density, whereas below it the perturbative hierarchy remains conventional.

Our results open several directions for future work. First, we have
restricted ourselves to a pure MD era ending at PBH evaporation; the
matching of the third-order SIGWs onto the subsequent radiation era,
including the transition dynamics studied at second order in
Refs.~\cite{Inomata:2019ivs,Inomata:2019zqy,Papanikolaou:2022chm,Kumar:2024hsi},
remains to be worked out. Since the secular growth continues until the
end of the MD era, the final spectrum depends sensitively on when the
transition happens, and the transition computation is likely to amplify
the effect found here. Second, the gauge dependence of the induced
gravitational waves, studied at second order in
Refs.~\cite{Hwang:2017oxa,Tomikawa:2019tvi,Inomata:2019yww,Domenech:2020xin,Ali:2020sfw},
deserves a dedicated analysis at third order; we have adopted the
Newton gauge throughout, where the perturbative expansion is known to
be well behaved at second order. Third, the results obtained in this work should be interpreted as constraints on the PBH-dominated era derived under the assumption that only second-order and third-order SIGWs are taken into account. The secular growth found here has a natural continuation to arbitrary orders: the vertex rules of
Ref.~\cite{Zhou:2024ncc} show that the tree-level contributions to the
$n$-th order SIGWs built from the frozen first-order scalars and the
growing lower-order modes inherit additional powers of $x$ at each
order, suggesting that the whole perturbative tower diverges during a
long MD era. If confirmed, this would indicate that an early MD era is
perturbatively unstable against the production of induced gravitational
waves, with profound implications for PBH domination scenarios and for
the use of SIGWs as probes of the small-scale primordial spectrum. We leave these questions to future work.

\appendix

\section{Wick contractions}\label{Ap:wick}

\begin{table}[t]
	\centering
	\begin{tabular}{cccc}
		\toprule
		contraction & $\pb'$ & $\qb'$ & $(u',\,v',\,\ubar',\,\vbar')$\\
		\midrule
		1 & $-\pb$ & $-\qb$ & $(u,\,v,\,\ubar,\,\vbar)$\\
		2 & $-\pb$ & $\qb-\pb$ & $(u,\,v,\,\vbar,\,\ubar)$\\
		3 & $\qb-\kb$ & $\qb-\pb$ &
		$(\vbar v,\,w,\,u/w,\,\ubar v/w)$\\
		4 & $\qb-\kb$ & $\pb-\kb$ &
		$(\vbar v,\,w,\,\ubar v/w,\,u/w)$\\
		5 & $\pb-\qb-\kb$ & $-\qb$ &
		$(\ubar v,\,\sigma,\,u/\sigma,\,\vbar v/\sigma)$\\
		6 & $\pb-\qb-\kb$ & $\pb-\kb$ &
		$(\ubar v,\,\sigma,\,\vbar v/\sigma,\,u/\sigma)$\\
		\bottomrule
	\end{tabular}
	\caption{The six cross Wick contractions of the six-point function
	$\langle\zeta_{\kb-\pb}\zeta_{\pb-\qb}\zeta_{\qb}
	\zeta_{\kb'-\pb'}\zeta_{\pb'-\qb'}\zeta_{\qb'}\rangle$. The last
	column gives the arguments of the primed kernel function
	$I^{(3)\prime}_{j}=I^{(3)}_{j}(u',v',\ubar',\vbar',x)$, with
	$u'=|\kb'-\pb'|/k$, $v'=p'/k$, $\ubar'=|\pb'-\qb'|/p'$,
	$\vbar'=q'/p'$, $\kb'=-\kb$, $w\equiv|\kb-\qb|/k$ and
	$\sigma\equiv\sqrt{1+u^{2}-w^{2}+v^{2}(\ubar^{2}+\vbar^{2}-1)}$.
	They were obtained by solving the momentum $\delta$ functions and
	verified numerically. Each contraction yields the same product of
	primordial spectra. Owing to the MD symmetry
	\eqref{eq:MDsym}, the primed kernel functions within each pair
	$\{1,2\}$, $\{3,4\}$, $\{5,6\}$ coincide, leading to the grouped
	kernel functions \eqref{eq:IMDa}. }
	\label{tab:wick}
\end{table}

The six cross contractions are listed in table~\ref{tab:wick}. Wick's theorem yields $15$ pairings in total. Among them, the $6$ cross contractions, in which each of the three $\zeta$ fields in $h^{\lambda,(3)}$ is contracted with one of the three $\zeta$ fields in $h^{\lambda',(3)}$, correspond to one-particle-irreducible loop diagrams. The remaining $9$ pairings correspond to one-particle-reducible diagrams. As shown rigorously using the vertex rules in Ref.~\cite{Zhou:2024ncc}, these reducible contributions vanish identically after performing the azimuthal integration over the loop momentum. Therefore, only the $6$ irreducible pairings contribute to the power spectrum for an arbitrary primordial spectrum.

\section{Momentum polynomials}\label{Ap:poly}

The momentum polynomials are computed from the compact definition
\eqref{eq:Pijdef}. We have the following relation:
\begin{equation}\label{eq:Pijproj}
	\Pb^{ij}
	=\widetilde{\Ssf}^{lm}_{i}\,\widetilde{\Ssf}^{j}_{lm}
	-\frac12\,\widetilde{\Ssf}_{i}\,\widetilde{\Ssf}_{j}\,,\qquad
	\widetilde{\Ssf}^{lm}_{i}\equiv
	P^{l}{}_{a}(\hat\kb)P^{m}{}_{b}(\hat\kb)\Ssf^{ab}_{i}\,,
	\qquad
	\widetilde{\Ssf}_{i}\equiv P_{lm}(\hat\kb)\Ssf^{lm}_{i}\,.
\end{equation}
All scalar products of the five vectors
$\{\hat\kb,\pb,\qb,\pb',\qb'\}$ reduce, through table~\ref{tab:wick}
and Eq.~\eqref{eq:angles}, to polynomials in $(u,v,w,\ubar,\vbar)$, so
every $\Pb^{ij}_{a}$ is a rational polynomial in these variables times
$k^{4}$. For example,
\begin{equation}\label{eq:P44ex}
	\Pb^{44}_{(1)}=\Pb^{44}_{(2)}
	=\frac12\,p^{4}\sin^{4}\theta_{p}
	=\frac{k^{4}v^{4}}{2}\left[1-\left(\frac{1+v^{2}-u^{2}}{2v}\right)
	^{2}\right]^{2}.
\end{equation}
The polynomials were computed symbolically and verified numerically at
multiple physical points against explicit polarization-tensor
contractions. In the monochromatic limit they reduce to the explicit
expressions listed in appendix~D of the RD reference
\cite{Zhou:2021vcw}. The entries
$\Pb^{13}_{a}=\Pb^{23}_{a}=\Pb^{43}_{a}=\Pb^{33}_{a}=0$ printed
there are confirmed. These identities hold only on the monochromatic support, as we now show.

\paragraph{Vector-induced identities on the monochromatic support.}
In Eq.~\eqref{eq:S3id}, the first identity follows from the definition in \eqref{eq:structs}: with the substitution $(\pb',\qb')=(-\pb,-\qb)$ the
structure is proportional to $\pb\cdot\qb\,[p_{l}\bar q_{m}+p_{m}\bar
q_{l}]$ with $\bar\qb=-P(\hat\pb)\qb$, while $(\pb',\qb')=(-\pb,\qb-\pb)$
gives the same expression with the opposite sign of
$P(\hat\pb)(\qb-\pb)=P(\hat\pb)\qb$; the two cancel because the
prefactors satisfy $2\,\pb\cdot\qb-p^{2}=q^{2}-|\pb-\qb|^{2}=0$ on the
monochromatic support $q=|\pb-\qb|=k_{*}$. For the second and third
pairs, the two transverse vectors obey
$P(\hat{\pb}')\qb'_{1}=-P(\hat{\pb}')\qb'_{2}$ because
$\qb'_{1}+\qb'_{2}=\pb'$, and the equal-magnitude conditions again make
the prefactors equal and opposite. These pointwise identities imply
$\Pb^{i3}_{a}=0$ for all $i$ and $a$, hence Eq.~\eqref{eq:h3zero}. For
a general primordial spectrum, the equal-magnitude conditions fail and
all sixteen terms contribute.

\section{Numerical implementation}\label{Ap:num}

The five-fold integral \eqref{eq:master} is evaluated on a
tensor-product grid after mapping each finite-range variable with a
midpoint rule; the $w$ integral is remapped with
$w^{2}=w_{-}^{2}+(w_{+}^{2}-w_{-}^{2})\sin^{2}\varphi$ to soften the
inverse-square-root endpoint singularities of the measure
\eqref{eq:measure}, and the upper limits of $v,\vbar$ are truncated at
the support of the primordial spectrum. For the monochromatic spectrum
the reduced two-fold form \eqref{eq:master-mono} is evaluated with
$200\times200$ points per wavenumber, and the
results are bitwise stable against doubling the resolution except in
the immediate vicinity of the kinematical thresholds.

For the broad PBH spectrum \eqref{eq:PPhiPBH}, the five-fold grid
suffers from aliasing of the inverse-square-root peaks of the measure,
so we evaluate the six-dimensional integral \eqref{eq:P6d} directly by
a Cartesian Monte Carlo integration over $(\pb,\qb)$, with $3\times10^{6}$ samples per point. The
two methods agree within the Monte Carlo errors at intermediate points;
for example, at $y_d=10$ the grid gives $1.36\times10^{-9}$ while the
Monte Carlo gives $(1.348\pm0.057)\times10^{-9}$ for
$\Psf^{(3)}_{h}$. The Monte Carlo error bars are displayed in
figure~\ref{fig:pbhspec}.

We performed the following self-consistency checks:
\begin{enumerate}
	\item the measure \eqref{eq:measure} reproduces
	$\int\dd^{3}p\,\dd^{3}q\,e^{-p^{2}-q^{2}}=\pi^{3}$ (with $k=1$)
	and $\int\dd\ubar\;(\text{measure})=\pi\vbar$ analytically, and
	passes independent Monte Carlo tests;
	\item the second-order pipeline reproduces the late-time analytic
	monochromatic MD spectrum, eq.~(40) of Ref.~\cite{Kohri:2018awv},
	in its domain of validity $x\gg1$, and the numerical second-order
	bound of Ref.~\cite{Papanikolaou:2020qtd} at the per-cent level;
	\item the exact symmetry $\Psf^{ij}_{h}=\Psf^{ji}_{h}$,
	Eq.~\eqref{eq:Psym}, is verified for all ordered pairs: the
	residuals are $2\times10^{-6}$ for $(1,2)$, $5\times10^{-5}$ for
	$(1,4)$ and $3\times10^{-4}$ for $(2,4)$ in relative terms, and
	shrink with increasing resolution;
	\item all pairs involving the vector-induced contribution
	vanish for the monochromatic spectrum: the pairs $(i,3)$ vanish
	pointwise at the integrand level, while the pairs $(3,i)$
	integrate to $10^{-13}$ of the total through the cancellation
	described in section~\ref{sec:h3zero}, providing an indirect but
	stringent test of the integration;
	\item the substitution rules for the kernel-function arguments of table~\ref{tab:wick} were
	verified numerically against direct evaluations of
	$(u',v',\ubar',\vbar')$ at random points of the integration domain.
\end{enumerate}

\acknowledgments
We thank Dr. Q.H. Zhu and Dr. X.K. Zhang for useful discussions.

\bibliography{biblio}

\end{document}